 \documentclass[aps,pre,reprint,showpacs,showkeys,noeprint,longbibliography]{revtex4-1}
\usepackage[utf8]{inputenc}
\usepackage{cmap} 
\usepackage[T1]{fontenc}
\usepackage{microtype}
\usepackage{amsmath,amssymb,mathtools}
\usepackage{txfonts}

\usepackage{mathrsfs}
\usepackage{graphicx,grffile,textcomp}
\graphicspath{.}
\usepackage[outdir=./]{epstopdf} 
\usepackage{textcomp}
\usepackage{booktabs,tabularx,dcolumn}
\newcolumntype{d}[1]{D{.}{.}{#1}}
\usepackage{titlesec}
\usepackage{url,hyperref}
\usepackage[usenames,dvipsnames,table]{xcolor}
\hypersetup{colorlinks=true, linkcolor=BrickRed, urlcolor=blue!50!black, citecolor=blue!50!black}

\begin{document}

\title{Effect of Cylindrical Confinement on the Collapse Dynamics of a Polymer}
\author{Shubham Thwal}\email[]{shubhamthwal1@gmail.com}
\affiliation{Amity Institute of Applied Sciences, Amity University Uttar Pradesh, Noida 201313,
India
}
\author{Suman Majumder}\email[]{smajumder@amity.edu,\ suman.jdv@gmail.com}

\affiliation{Amity Institute of Applied Sciences, Amity University Uttar Pradesh, Noida 201313,
India
}
\begin{abstract}
\noindent\textbf{Abstract—}
 Structure and dynamics of a polymer under confinement gets significantly altered due to the imposed  geometric restrictions. Using molecular dynamics simulations, here, we explore the effect of cylindrical confinement on the kinetics of collapse of a homopolymer, when the solvent condition is abruptly changed from good to poor. The observed phenomenology for a range of the cylinder radius $R$, reveals two distinct stages of the collapse. The first stage is highlighted by the formation and growth of local connected  clusters resembling a pearl necklace, eventually ending with a single sausage-like cluster. In the second stage, the sausage-like intermediate approaches a spherical globule via surface-energy minimization. These two stages are disentangled using a shape parameter of the individual pearls or clusters, allowing us to also extract the respective relaxation times, and thereby their scaling behaviors with respect to the length of the polymer. We find that the pearl-necklace relaxation time $\tau_p$ is independent of $R$. On the other hand, the sausage-relaxation time $\tau_s$ varies inversely up to a certain $R$, beyond which it also saturates. From the Arrhenius plots of the temperature dependence of $\tau_p$ and $\tau_s$, we extract the activation energies  $E_{\rm a}$ of the two stages. While the estimated $E_{\rm a}$ for the pearl-necklace stage is independent of $R$,  for the sausage relaxation it is significantly higher in the strongly confined case than in the weakly one. Surprisingly, at a fixed temperature, the growth of the average cluster size obeys a universal power law irrespective of $R$. However, for a fixed $R$, the behavior is rather non-universal with respect to temperature. We propose viable scenarios for experimental realization of polymer collapse inside cylindrical nanochannels.

\end{abstract}

\maketitle

\section{Introduction}\label{intro}
The coil-globule or collapse transition is the most basic and widely investigated conformational transformation of a polymer. There, typically, an extended polymer in good solvent, collapses into a compact structure on changing the solvent condition to poor or unfavorable \cite{flory1953principles,rubenstein2003}. Such a collapse of the polymeric backbone of a protein is also an integral part of its overall folding process \cite{camacho1993kinetics,pollack2001time,sadqi2003fast,haran2012,udgaonkar2013,reddy2017collapse}. Hence, over the years study of the collapse kinetics of a polymer has received a lot of attention \cite{de1985kinetics,byrne1995kinetics,kuznetsov1996kinetic,halperin2000early,pitard2001glassy,dokholyan2002,abrams2002collapse,kikuchi2005kinetics,xu2006first,ye2007many,majumder2015cluster,majumder2016evidence,majumder2017kinetics,christiansen2017coarsening,majumder2019pearl,majumder2020understanding,schneider2020,majumder2024temperature,majumder2025scale} with the motivation to gain any universal insights, which might be applicable to the folding kinetics of a protein as well.  
\par
Most of the above mentioned studies on polymer collapse were carried out in a bulk solution or unconfined environment, free from any geometric restrictions. However, realistic conditions usually are associated with some form of geometrical confinement, which have a huge impact on both the conformation and dynamics of a polymer \cite{de1979scaling,milchev2011,reisner2012dna,wang201750}. Examples of such systems include biopolymers confined in cellular environments and the synthetic polymers in microfluidic systems \cite{perkins1997single,kindt2001dna,purohit2003,lu2005folding,squires2005,jun2006entropy,persson2010dna}. In this context, the cylindrical confinement is among the widely encountered ones, e.g., rod-shaped bacterial cells, viral capsids, ribosomal exit tunnels and protein-conducting channels. Restricted transverse fluctuations under such conditions strongly affect the dynamic pathways of the associated conformational transitions. For instance, DNA condensation, packaging and ejection in bacteriophages occurring under extreme confinement shapes its biological function \cite{purohit2003}. Similarly, chromosomal organization and segregation within rod-shaped bacterial cells can be understood using modified polymer physics laws under cylindrical confinement \cite{jun2006entropy}. Proteins translocating through ribosomal exit tunnels also  experience quasi-cylindrical confinement that suppresses its folding by altering the collapse pathways relative to what is expected in a bulk solution \cite{lu2005folding}. Understanding how a polymer collapses from an extended to a compact state in a cylindrical confinement, therefore, has a strong biological relevance. In fact the effect of any such geometrical confinement on the collapse dynamics has rarely been studied \cite{das2010effect}.
\par
Despite the above mentioned importance, the kinetics of collapse of a flexible homopolymer inside a cylindrical confinement remains completely unexplored. For collapse of a polymer in a bulk solution, de Gennes proposed the first phenomenological picture. It suggests the formation of a sausage-like intermediate before the polymer achieves a spherical globular state \cite{de1985kinetics}. Later, Halperin and Goldbart (HG) introduced the pearl-necklace model proposing a multi-step collapse process \cite{halperin2000early}. According to HG the collapse begins with the  formation of local pearls or clusters connected via bridges of monomers, resembling a necklace of pearls. These pearls subsequently grow via merging with each other, and eventually yielding a single spherical globule. In the past, most of the simulation studies as well as experiments observed such a multi-stage collapse with pearl-necklace intermediates \cite{abrams2002collapse,kikuchi2005kinetics,xu2006first,ye2007many,majumder2015cluster,majumder2016evidence,majumder2017kinetics,christiansen2017coarsening,majumder2019pearl,majumder2020understanding,schneider2020}. Lately, it has been demonstrated that depending on the solvent viscosity and temperature one can also have a combination of the pearl-necklace and sausage-like intermediates \cite{majumder2024temperature}.  

\par
Traditionally, the kinetics of collapse in a bulk solution is probed via the following two nonequilibrium scaling laws -- (i) dependence of the overall collapse time $\tau_c$ with the  length of the polymer chain, typically measured in terms of the number of monomers $N$ and (ii) growth of the average cluster size $C_s$ with time $t$. There are also studies that quantify a two-time quantity to characterize the associated physical aging and related scaling \cite{pitard2001glassy,dokholyan2002,majumder2016evidence,majumder2016aging,majumder2017kinetics,christiansen2017coarsening,majumder2018scaling}. Since the collapse is a multi-stage process, naturally, a deeper  understanding of the overall process demands disentanglement of the time scales associated with these stages. Theoretical predictions on the scaling of the time scales were in place for long \cite{halperin2000early,klushin1998}. However, due to technical difficulties in extraction of the time scales from simulation data, understanding of their scaling behavior at various stages of the collapse is still developing  \cite{schneider2020,majumder2024temperature}.

\par
Here, we study the effect of cylindrical confinement on the collapse dynamics of a single flexible homopolymer using molecular dynamics (MD) simulations. By varying the strength of confinement and temperature, we focus on analyzing how the phenomenology of the collapse and associated relaxation times get modified due to the imposed geometric restrictions. Our results suggest that the collapse occurs in two stages, characterized by the pearl-necklace phenomenology at the beginning followed by the sausage relaxation. Based on this picture we disentangle the relaxation times concerning the two stages and their respective scaling behaviors with respect to the polymer length. From the temperature dependence of the relaxation times and corresponding Arrhenius behaviors we also gauge an idea about the activation energies associated with the two stages. We also find that the cluster growth dynamics in the pearl-necklace stage shows a universal behavior irrespective of the confinement strength, however, it depends strongly on the temperature.

\par
\par
The rest of the paper is organized in the following order. In Sec.\ \ref{methods}, we present details of the polymer model and confinement, along with a description of the MD simulations performed. Following that in Sec.\ \ref{results}, we present our results. Finally, we conclude in Sec.\ \ref{conclusion} with a summary of the main results and a future outlook.

\section{Model and Method}\label{methods}

We consider a coarse-grained bead-spring model of a linear homopolymer chain made of $N$ monomers (or beads) with uniform mass $m$ and diameter $\sigma$. The bonded interaction acting only between two successive monomers at a distance $r$ apart is given by the standard finitely extensible non-linear elastic (FENE) potential \cite{kremer1990dynamics}

\begin{align}\label{fene_potential}
V_{\text{FENE}}(r) &= -\frac{1}{2} K R_0^2 \ln\left(1 - \left(\frac{r}{R_0}\right)^2\right); \quad r < R_0,
\end{align}
where $K=30\varepsilon/\sigma^2$ and $R_0=1.5\sigma$ represent, respectively, the spring constant and maximum extension of the bond. The nonbonded interaction between two monomers is modeled via the standard Lenard-Jones (LJ) potential
\begin{align}\label{lj_potential}
V_{\rm LJ}(r) = 4 \varepsilon \left[ \left( \frac{\sigma}{r} \right)^{12} - \left( \frac{\sigma}{r} \right)^6 \right],
\end{align}
where $\varepsilon=1$ sets the interaction strength. The volume exclusion between the monomers is maintained via the repulsive part of $V_{\rm LJ}$, and the collapse transition of the polymer as a function of temperature is ensured by the attractive part. For computational efficiency, instead of using the full LJ potential, we truncate and shift it at $r_c=2.5\sigma$ to get
\begin{align}\label{kremer_lj_potential}
V_{\rm LJ}^{\rm mod}(r) =
\begin{cases}
V_{\rm LJ}(r) - V_{\text{LJ}}(r_c) ; & r < r_c \\[8pt]
0; & r \ge r_c.
\end{cases}
\end{align}
\par
 We consider the polymer to be confined inside a cylinder of finite radius $R$ and infinite length. The wall of the cylinder is considered to be rigid. The monomers of the polymer interact with the cylinder wall in a purely repulsively manner. This is achieved by using the same LJ potential described in Eq.\ \eqref{lj_potential} as a function of the radial distance of the monomers from the wall, and correspondingly using $r_c = 2^{1/6} \sigma$ in Eq.\ \eqref{kremer_lj_potential}.
\par
Using the set up described above, MD simulations are performed at constant temperature, employing the standard velocity-Verlet scheme to integrate the equations of motion of the polymer beads \cite{frenkel}. The temperature is kept under control using the Nos\'e-Hoover (NH) chain thermostat \cite{nose1984unified,hoover1985canonical,martyna1992nose}. Since NH thermostat maintains the conservation of linear momentum, it has been used to realize hydrodynamic regime in fluid systems \cite{majumder2013effects}. In our simulations ${\varepsilon}/{k_B}$ is the unit of temperature and $\sigma$ is the unit of length. We choose the discrete time step of integration $\Delta t = 0.005\ \tau_0$, where the standard LJ unit of time is $\tau_0 = \sqrt{{m \sigma^{2}}/{\varepsilon}}$. For convenience, we set both $m=1$ and $\sigma=1$. 
\par
\begin{figure*}[t!]
\centering
\includegraphics[width=0.98\textwidth]{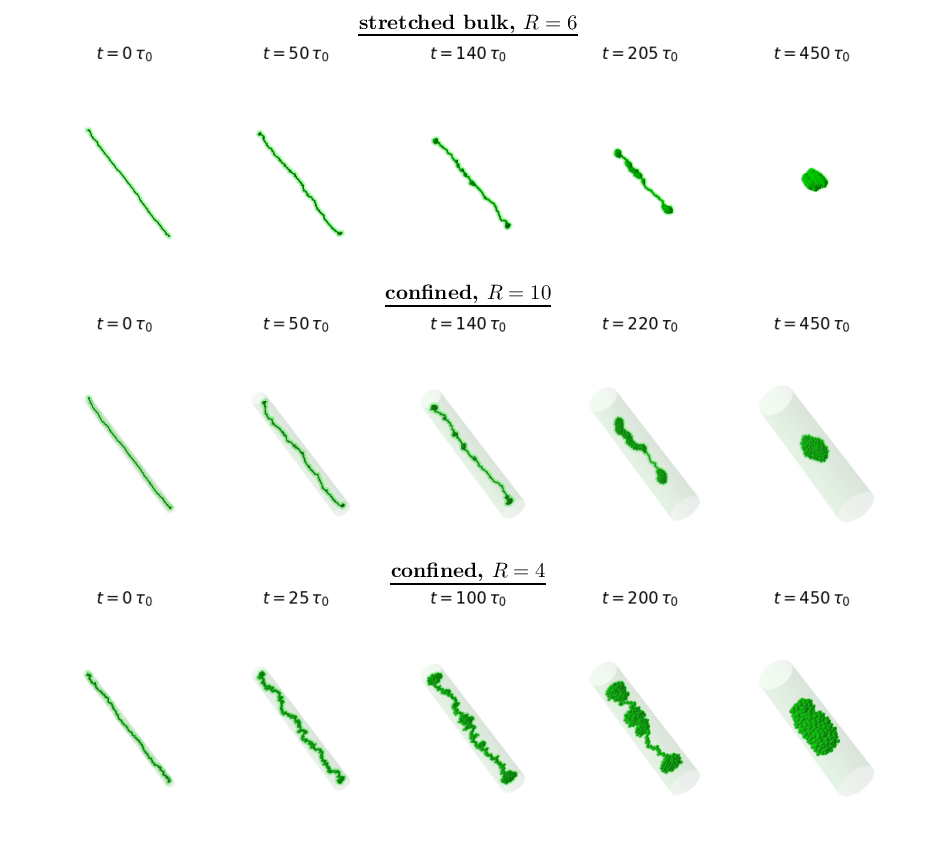}
\caption{Time-evolution snapshots of a polymer of length $N=1024$, undergoing collapse transition at temperature $T=1.0$, for three different cases -- a stretched bulk, and inside cylindrical confinement of radius $R=10$ and $4$. Here, and for all the subsequent plots concerning the stretched bulk case, $R$  refers to the radius of the cylinder where the equilibration at high $T$ is performed, before getting rid of the confinement for the quench to a low $T$. In the confined  cases only the relevant portion of the infinitely long cylinders are shown. To obtain a clear view of the temporal changes of the polymer conformation, the appearances of the cylinder at different times are adjusted accordingly.}
\label{fig:snapshots}
\end{figure*}

The initial condition is prepared by equilibrating the polymer inside the cylinder of choice at a temperature $T=5$ yielding an elongated chain conformation (see Fig.\ \ref{fig:snapshots}). The temperature is then set to a much lower value ensuring the collapse of the polymer. For faster computation, we make use of the LAMMPS simulation package \cite{plimpton1995}.
As per our requirement, we perform simulations of different chain lengths $N \in [256,1024]$ at different temperatures $T\in [0.3,1.5]$, for a range of cylinder radius $R \in [3,10]$. All subsequent results (except the time-evolution snapshots) are averaged over a minimum of $50$ independent realizations. 

\par 
The results from simulations of polymers under confinement, at times are compared with results from  the collapse dynamics of a polymer in a bulk solution, i.e, in absence of any confinement. We refer to such a case as usual bulk. Note that in the usual bulk condition the initial conformation of the polymer is an extended coil, mimicking a polymer in a good solvent. Apart from that, in order to explicitly gauge the confinement effect we also present results from a special case, which we refer to as stretched bulk. In that case, the initial conformation of the polymer is also a stretched or elongated one, generated inside the cylindrical confinement. Afterward, when the temperature quench is performed we allow it to evolve in absence of any confinement, i.e., in bulk condition.
%

\section{Results}\label{results}

\subsection{Relaxation Dynamics}

We begin by showing the time-evolution snapshots of a polymer during its collapse in Fig.\ref{fig:snapshots}. Results for three different cases, as mentioned, are presented.  In case of the stretched bulk, it can be observed that initially clusters are formed at the ends of the chain (snapshot at $t=50\ \tau_0$). This is in contrast with the pearl-necklace picture observed in the usual bulk case, where cluster nucleation occurs uniformly along the chain \cite{halperin2000early,majumder2017kinetics,majumder2025scale}. After a brief period, clusters start nucleating uniformly along the contour of the polymer (snapshot at $t=140\ \tau_0$). Following that, the collapse enters a coarsening regime which proceeds via growth and coalescence of clusters eventually leading to the formation of a single cluster. Later, this single cluster transforms to a compact spherical globule. Similar sequence of events till the formation of a single cluster unfolds for polymers confined inside cylinders as well. The final conformation (or at long time) depends on the strength of the confinement, i.e., on $R$. The shape of the polymer after it becomes a single cluster is cylindrical, which via minimization of surface energy transforms to a spherical globule for the case with a larger $R$. This is reminiscent of de Gennes' sausage picture \cite{de1985kinetics}. For smaller $R$, however, it cannot completely relax to a spherical globule. From an inspection of the collapse pathways for different $R$ (all are not shown here) for a fixed $N$, we notice that as $R$ decreases, the final stage of transformation of a cylindrical globule to a spherical one slows down. Overall, it can be concluded that depending on $R$, there are two distinct stages of collapse, respectively, highlighted by a pearl-necklace intermediate followed by a sausage-like intermediate. Such a combination of both the intermediates has been encountered for a polymer collapsing in a viscous solvent at low $T$ \cite{majumder2024temperature}. 
\par
\begin{figure}[t!]
\centering
\includegraphics[width=0.48\textwidth]{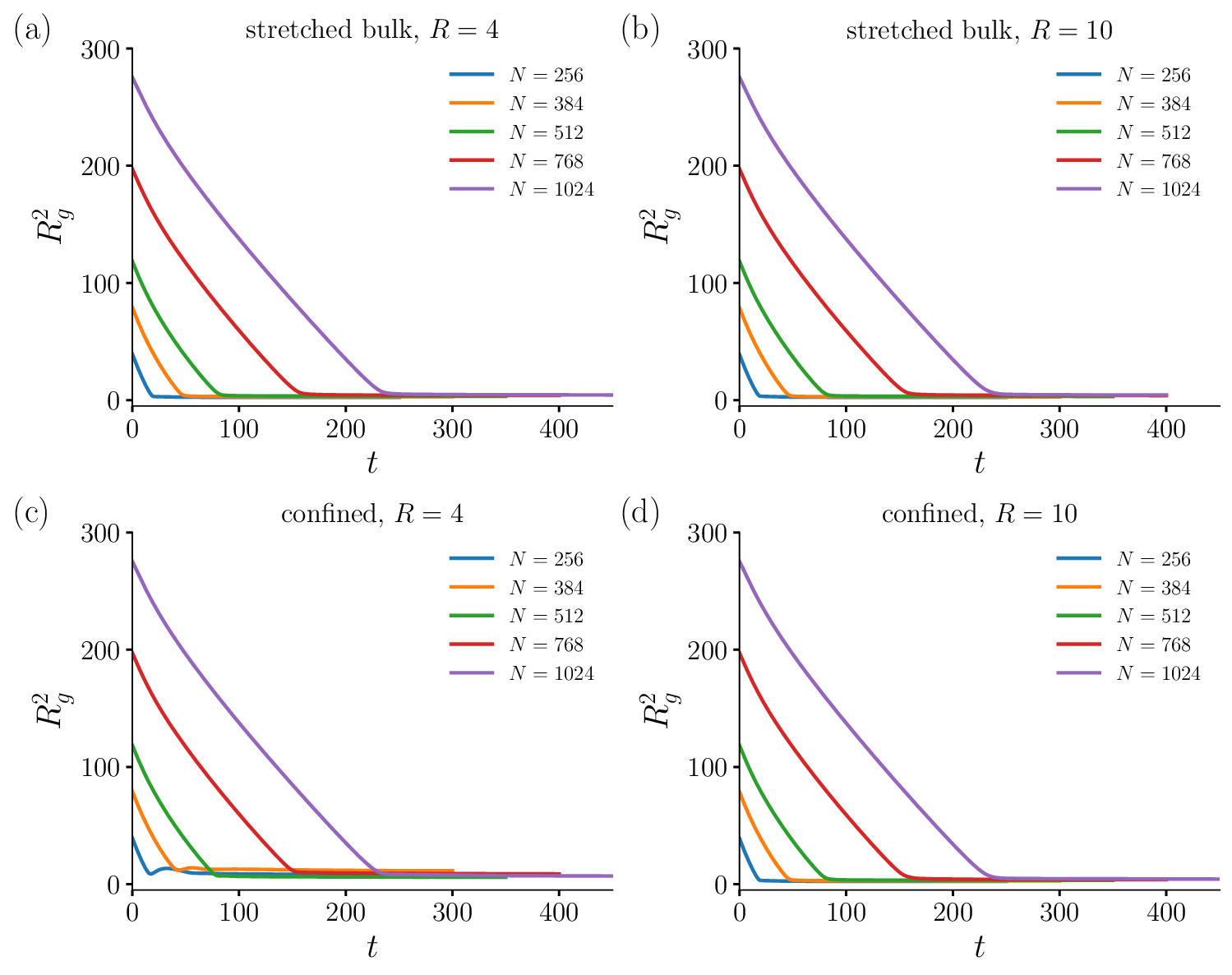}
\caption{Time dependence of the squared radius of gyration $R_g^2$ of a collapsing polymer at a temperature $T=1.0$, for different cases in (a)-(d), as mentioned. In each case, results for different chain lengths $N$ are presented.}
\label{fig:gyration}
\end{figure}

Next, we move on to quantitative analyses of the two stages. At first, following the traditional approach we calculate the squared radius of gyration of the polymer 

\begin{equation}
 R_g^2 = \left \langle \frac{1}{2N^2} \sum_{i=1}^{N} \sum_{j=1}^{N} \left( \mathbf{r}_i - \mathbf{r}_j \right)^2 \right \rangle,
\end{equation}
where $\mathbf{r}_i$ is the position vector of the $i$-th monomer. Unless otherwise mentioned, now onward $\langle \dots \rangle$ denotes average over many independent initial realizations. The time dependence of $R_g^2$ during collapse of polymers  under different conditions at $T=1.0$ is presented in Fig.\ \ref{fig:gyration}.  In each case, we show results for five different $N$. For a fixed $N$, the decay of $R_g^2$ shows similar behavior for all the cases. Clearly, as $N$ increases the decay to its final value is slower, indicating an $N$-dependent overall collapse time. For all the cases, the data do not show any signature of the intermediate stages of the collapse. Certainly, one could use the data to extract an overall collapse time. Since our focus is on disentangling the time scales associated with the two stages of collapse, here, we do not extract the overall collapse time.

\par

\begin{figure}[t!]
\centering
\includegraphics[width=0.48\textwidth]{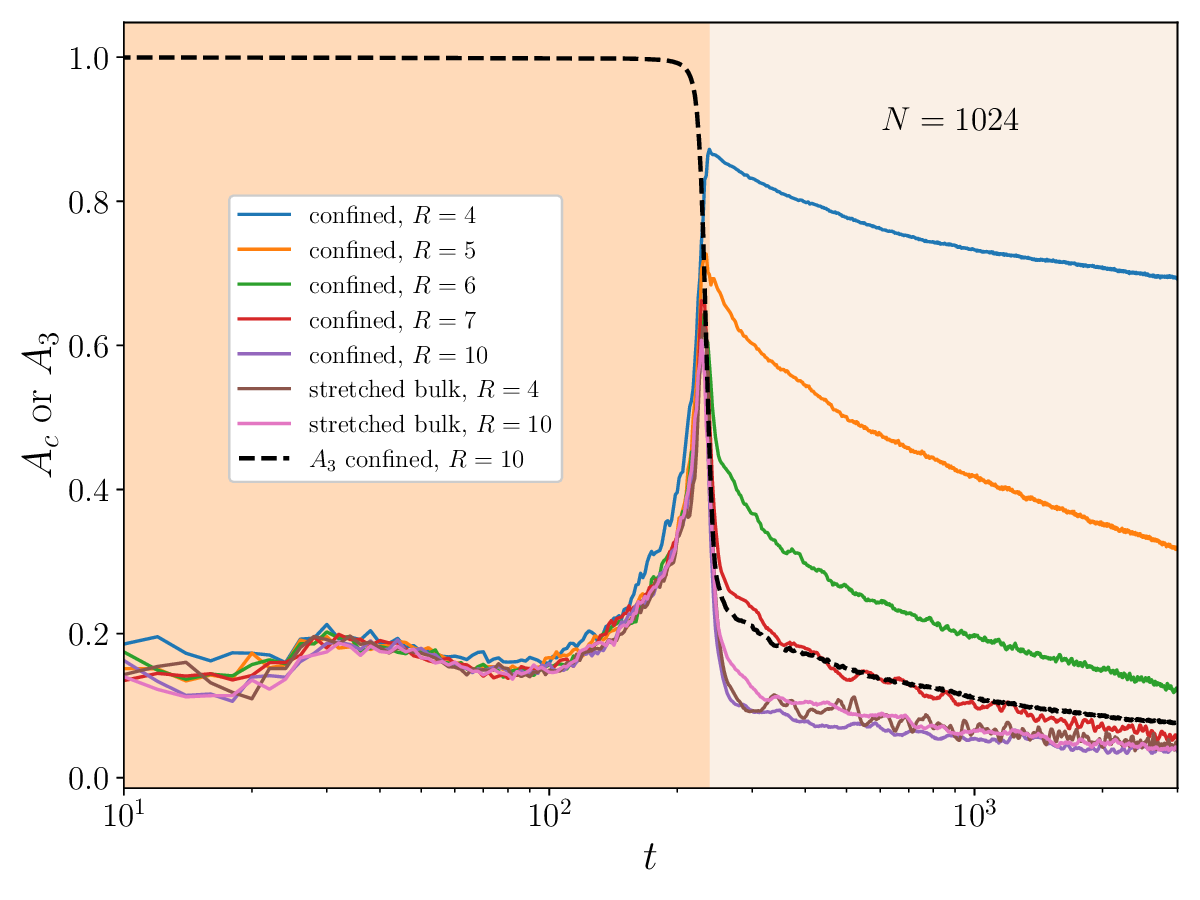}
\caption{Time dependence of the average asphericity $A_c$ of individual clusters at a temperature $T=1.0$, for a collapsing polymer of length $N=1024$, under different conditions. Corresponding data (black dashed line) for the overall asphericity $A_3$ of a polymer under confinement of $R=10$ is also presented. The shades with different colors represent two stages of the collapse.}
\label{fig:asphericity}
\end{figure}

In Ref.\ \cite{majumder2024temperature}, it has been demonstrated that monitoring time dependence of shape factors provides precise information about the conformational changes of a collapsing polymer. Shape factors of the full polymer give an idea about the global geometry, averaged over all monomers. So, it is not expected that the local structural rearrangements during the collapse can be captured by it.  Nevertheless, the corresponding data of the overall asphericity $A_3$ of a polymer collapsing inside a cylinder with $R=10$ are presented in Fig.\ \ref{fig:asphericity}. There, it can be observed that during the pearl-necklace stage $A_3$ does not show any feature, and remains almost constant around unity, courtesy the cylindrical confinement. Afterward $A_3$ falls sharply marking the beginning of the sausage-relaxation stage.  This makes it a suitable candidate for identifying the crossover of the two stages, which can be obtained from the inflection point where $A_3$ starts to fall off. 
\par
Considering the formation of local clusters or pearls during the pearl-necklace stage, we opt for estimating the asphericity of these clusters as well. To proceed with that, we first identify the discrete clusters and count their total numbers $n_c(t)$ at a given time, following the prescription mentioned in Refs.\ \cite{majumder2020understanding,majumder2024temperature}. Once they are identified, using the coordinates of the monomers belonging to the  $k$-th cluster, we calculate the corresponding gyration tensor $\mathbf{Q}$ with elements

\begin{equation}\label{tensor}
 Q_{ij} = \sum_{l=1}^{m_k} (x_l^i-x_{\rm CM}^i)(x_l^j-x_{\rm CM}^j),~i,j=1,2,3,
\end{equation}
where $m_k$ is the number of monomers in the $k$-th cluster and $x_{\rm CM}^i$ represents the $i$-th component of the cluster's center-of-mass (CM) vector 
\begin{equation}\label{cluster_cm}
 \vec{r}_{\rm CM} = \frac{1}{N} \sum_{l=1}^{m_k} \vec{r}_i.
\end{equation}
From the eigenvalues $\lambda_i$ of $\mathbf{Q}$, we estimate the average asphericity of the clusters
\begin{equation}\label{asphericty}
 A_c =\left \langle \frac{1}{n_c(t)}\sum_{j=1}^{n_c(t)}\frac{1}{6}\sum_{i=1}^3\frac{(\lambda_i-\bar{\lambda})^2}{\bar{\lambda}^2}\right\rangle, 
\end{equation}
where $\bar{\lambda}$ is the mean eigenvalue. The measurement of $A_c$ tells how much the clusters deviate from a spherical symmetry. By definition, the value of $A_c$ lies between $0$ and $1$. The case with $A_c=0$ refers to a perfect sphere and $A_c=1$ indicates a perfectly cylindrical cluster.

\par
In Fig.\ \ref{fig:asphericity} we present the variation of $A_c$ of a polymer with time, during collapse under different conditions. The hump in the data at early time is due to the nucleation of clusters at the ends of the polymer, which we have also pointed out while presenting the time evolution snapshots in Fig.\ \ref{fig:snapshots}. Disregarding the initial hump, the plots clearly highlight the two distinct kinetic regimes. The first stage is the pearl-necklace stage, where $A_c$ increases to a maximum value, implying that the polymer has attained a cylindrical or sausage-like conformation. This trend is true for all the cases including the stretched bulk. It can also be noticed that the time when $A_c$ reaches its maximum, is independent of $R$. The concerned time essentially marks the end of the pearl-necklace stage and the beginning of the sausage-relaxation stage. For the stretched cases too, the data attain the maximum almost at the same time. In the second stage, i.e., during the sausage relaxation, $A_c$ decays from its maximum value toward a much smaller value corresponding to a more spherical shape. In this period, as $R$ increases the relaxation of $A_c$ occurs faster, indicating a strong dependence of the sausage relaxation on the confinement. In the limit of large $R$, the decay of $A_c$ becomes independent of $R$, as the sausage has spatial freedom to reorganize itself into a more stable spherical conformation. For the stretched-bulk case the dependence of the sausage relaxation on $R$ of the cylinder where the initial condition was prepared, is rather weak.

\par
The pearl-necklace stage is characterized by the formation of multiple compact and elongated clusters. In this stage, as already pointed out, there is a sharp growth in $A_c$, which stops after reaching its peak value $A_c^{\rm max}$. The corresponding time is defined as the pearl-necklace relaxation time $\tau_p$. For the quantification of the sausage-relaxation time, we first calculate 
\begin{equation}\label{Ac_max}
\Delta_{A_c}= A_{c}^{\max} - A_{c}(t \to \infty), 
\end{equation}
where $A_c$ in the long-time limit is obtained from the value achieved at the latest time in respective simulations. Using $\Delta_{A_c}$ we define an overall collapse time $\tau_c$ such that 
\begin{equation}\label{tau_c}
 A_c(t=\tau_c) = A_{c}(t \to \infty) + 0.5\Delta_{A_c}.
\end{equation}
Once both $\tau_p$ and $\tau_c$ are obtained, one can extract a timescale associated only with the relaxation of the sausage as 
\begin{equation}\label{tau_s}
 \tau_s= \tau_c-\tau_p.
\end{equation}

\begin{figure}[t!]
\centering
\includegraphics[width=0.48\textwidth]{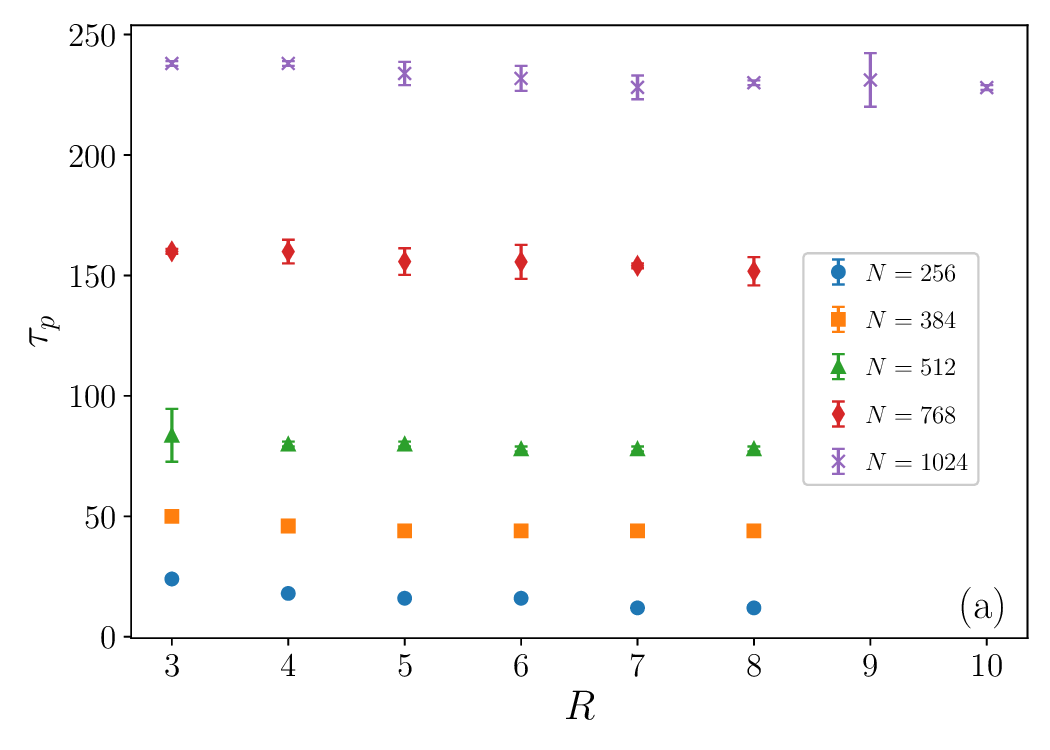}\\
\includegraphics[width=0.48\textwidth]{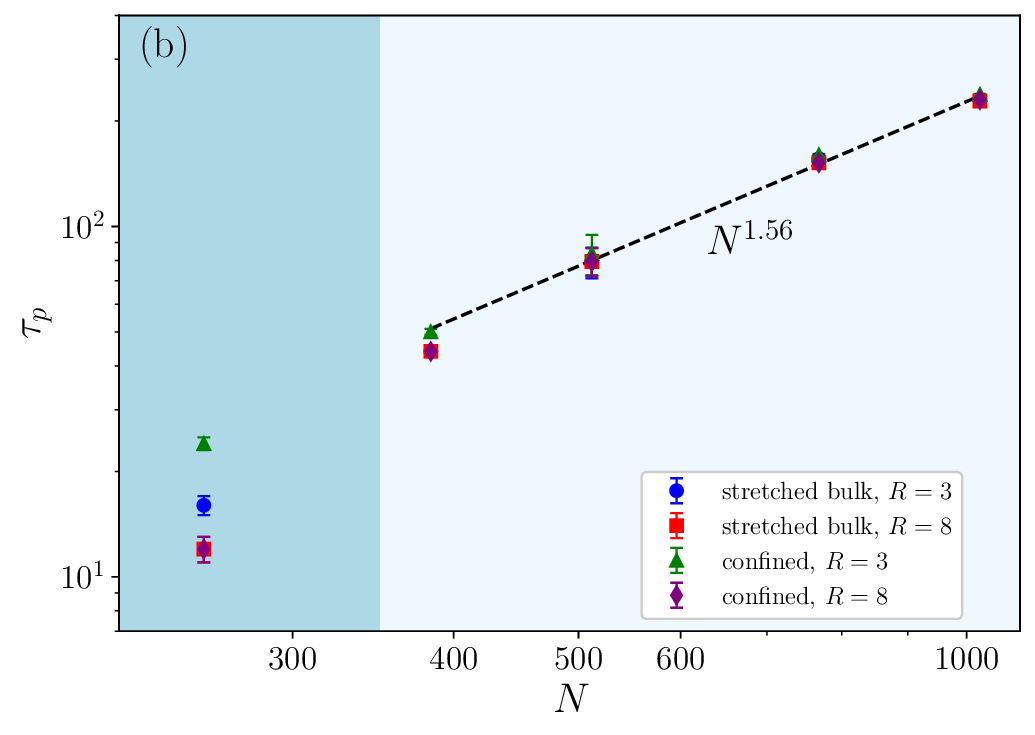}\\
\caption{(a) Dependence of the pearl-necklace relaxation time $\tau_p$ on the radius $R$ of the cylindrical confinement for polymers of different chain lengths $N$, at a fixed temperature $T=1.0$. (b) Scaling of $\tau_p$ with $N$, for both the confined and stretched-bulk polymer at two values of $R$. The dashed line represents the power-law scaling of the form  in Eq.\ \eqref{equation:tau_p_scaling} with $z_p=1.56$. The region shaded with lighter color marks the fitting range.}
\label{fig:tau_p_plot}
\end{figure}
\par
As already intuited, the plots of $\tau_p$ as a function of $R$ for the confined case presented in Fig.\ \ref{fig:tau_p_plot}(a) confirm that the relaxation dynamics in the pearl-necklace stage is largely independent of $R$. Similar behavior is also obtained for the stretched-bulk case as well (not shown here). Next, in Fig \ref{fig:tau_p_plot}(b) we demonstrate the scaling of $\tau_p$ with $N$ by plotting the data for both the confined and stretched-bulk polymer at two values of $R$. In all the cases, a linear behavior on a double-log scale confirms a power-law scaling, which we quantify using the ansatz \cite{majumder2024temperature}
\begin{equation} \label{equation:tau_p_scaling}
 \tau_p = \tau_p^{0} N^{z_p},
\end{equation}
where $\tau_p^{0}$ is the amplitude and $z_p$ is the corresponding dynamic exponent. Results from fitting of the ansatz in Eq.\ \eqref{equation:tau_p_scaling} using the data of $\tau_p$ for $N>256$ is presented in Table\ \ref{tab:tau_p_table_N}. As apparent from the similar behavior of different cases in Fig.\ \ref{fig:tau_p_plot}(b), the obtained range of values of $z_p \in [1.52,1.59]$ is quite narrow, and they yield an average $z_p \approx 1.56$. This can also be appreciated from the consistency of the data for all the cases with the dashed line representing $\tau_p \sim N^{1.56}$ in  Fig.\ \ref{fig:tau_p_plot}(b). So, it can be concluded that the scaling of the pearl-necklace relaxation time with respect to the chain length is largely independent of the confinement effect.

\begin{table}[b!]
\centering
\caption{Results from fitting of $\tau_p$ as a function of $N$ at $T=1.0$ for different cases using the ansatz in Eq.\ \eqref{equation:tau_p_scaling}. Note that we exclude the data for $N=256$ to avoid finite-size effects.}
\label{tab:tau_p_table_N}
\begin{tabular}{lll} 
\hline
\hline
~~~~~~Case & ~~~~~~$\tau_p^0$ &~~~~~$z_p$ \\
\hline
confined, $R=3$ & 0.0064(33) & 1.52(07) \\
confined, $R=8$ & 0.0036(22) & 1.59(09) \\
stretched bulk, $R=3$ & 0.0046(38) &1.56(12) \\
stretched bulk, $R=8$ & 0.0040(27) & 1.58(09) \\
\hline
\end{tabular}
\end{table}

\begin{figure}[t!]
\centering
\includegraphics[width=0.48\textwidth]{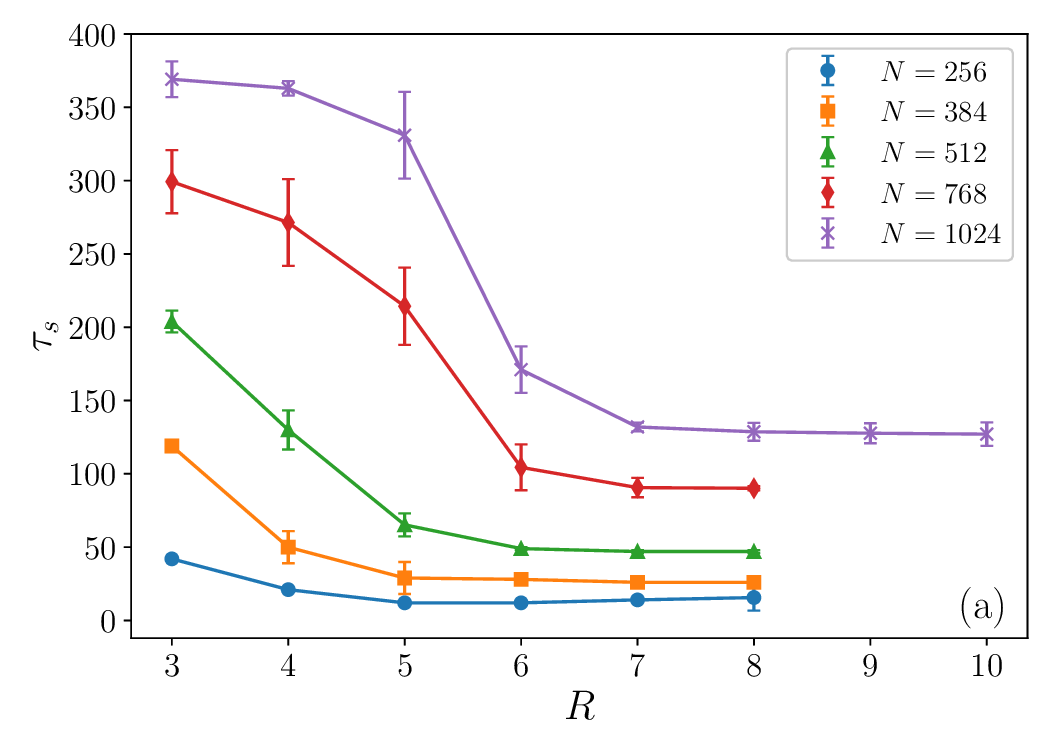}\\
\includegraphics[width=0.48\textwidth]{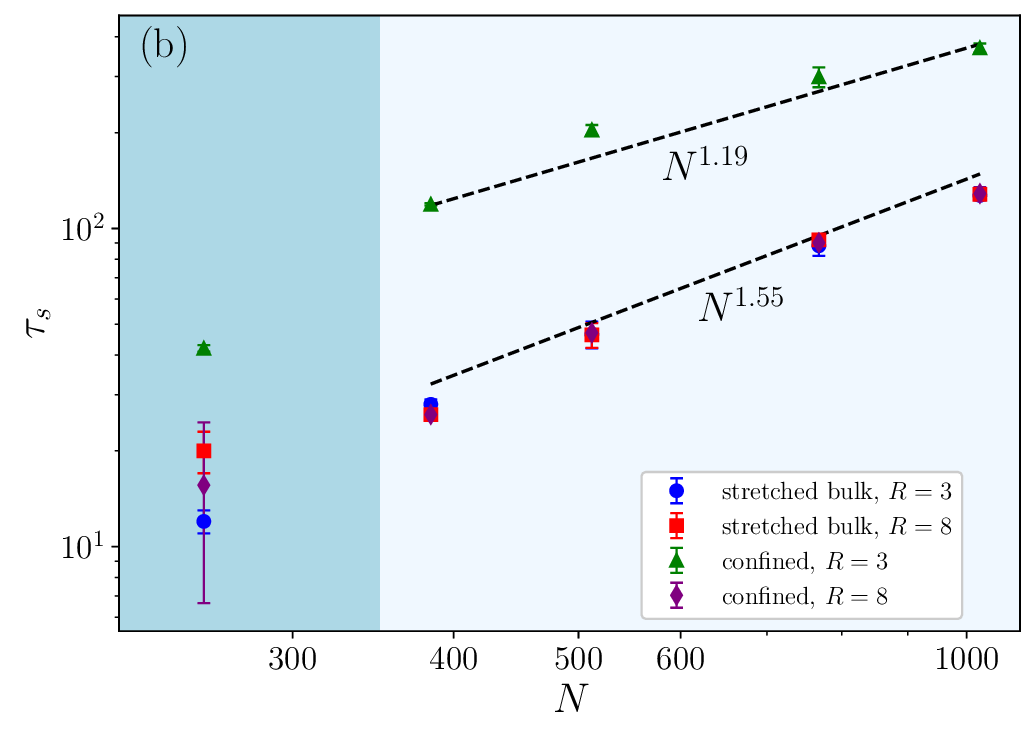}
\caption{(a) Dependence of the sausage relaxation time $\tau_s$ on the cylinder radius $R$ for confined polymers of different chain lengths $N$, at a fixed temperature $T=1.0$. (b) Scaling of $\tau_s$ with $N$, for both the confined and stretched bulk cases at two values of $R$. The dashed lines represent the power-law scaling of the form  in Eq.\ \eqref{equation:tau_s_scaling} with the quoted values of $z_s$. The region shaded with lighter color marks the fitting range.}
\label{fig:tau_s_plot}
\end{figure}
\par
 In contrast to the pearl-necklace relaxation, there is a strong effect of confinement on the dynamics of the sausage-relaxation stage, as demonstrated in Fig.\ \ref{fig:tau_s_plot}(a), showing plots of $\tau_s$ against $R$ for different $N$. The data show a decrease in $\tau_s$ with increasing $R$, implying a faster sausage relaxation as confinement loosens. This decrease in $\tau_s$ occurs until a certain $R$, beyond which it saturates. Note that the saturation of $\tau_s$ starts appearing at larger $R$ as $N$ increases. For the stretched-bulk case (not shown here) with fixed $N$, the behavior of $\tau_s$ is independent of $R$ of the cylinder where the initial conformation is prepared. 
 \par
 Next, in Fig.\ \ref{fig:tau_s_plot}(b) we show the $N$-dependence of $\tau_s$  for both the confined and stretched-bulk cases with $R=3$ and $8$. The choice of these two values of $R$ is motivated by the behavior observed in Fig.\ \ref{fig:tau_s_plot}(a). The choice $R=3$ ensures that one is in the $R$-dependent regime of $\tau_s$ for all $N$. On the other hand, the choice $R=8$ is a representative of the regime where $\tau_s$ is independent of $R$ for all $N$. For all the cases in Fig.\ \ref{fig:tau_s_plot}(b), one observes a linear behavior on a double-log scale, confirming a power-law scaling of $\tau_s$ with $N$. Except for the confined case with $R=3$, all the cases show almost identical variation of $\tau_s$ with $N$. It can be noticed that for the case of confined polymer with $R=3$, the values of $\tau_s$ for a fixed $N$ are significantly larger than the other cases. To further explore the difference in dynamics  in the two regimes of $R$, we quantify the power-law behavior of $\tau_s$ with $N$ using the ansatz
 \begin{equation}\label{equation:tau_s_scaling}
 \tau_s = \tau_s^{0} N^{z_s}, 
\end{equation}
where $\tau_s^0$ is the amplitude and $z_s$ is the associated dynamic exponent. Results of fitting this ansatz with the corresponding data are tabulated in Table\ \ref{tab:tau_s_table_N}. The value of $z_s$ for the confined case with $R=3$ is significantly smaller than the other cases. This implies a weaker dependence of $\tau_s$ on $N$ under strong confinement. Thus, the significantly larger value of $\tau_s$ than the other cases for a fixed $N$, is attributed to a much larger value of $\tau_s^0$ for the confined $R=3$ case. As the kinetics of the sausage-relaxation stage is supposed to depend on the rate at which the surface energy is minimized, we speculate that the physical significance of the fitting parameter $\tau_s^0$ is plausibly embedded in the associated surface free-energy barrier. The other three cases show a fair level of agreement among each other in terms of the values of $z_s$, with relatively large error bars. This can also be appreciated from the consistency of the data with the dashed line representing the power law $\tau_s\sim N^{1.55}$ in Fig.\ \ref{fig:tau_s_plot}(b). The value $1.55$ is the mean of $z_s$ quoted for the last three cases in Table\ \ref{tab:tau_s_table_N}.
\begin{table}[t!]
\centering
\caption{Results from fitting of $\tau_s$ as a function of $N$ at $T=1.0$ for different cases using the ansatz in Eq.\ \eqref{equation:tau_s_scaling}. Here also, we exclude the data for $N=256$.}
\label{tab:tau_s_table_N}
\begin{tabular}{lll} 
\hline
\hline
~~~~~~Case & ~~~~~~$\tau_s^0$ &~~~~~$z_s$ \\
\hline
confined, $R=3$ & 0.0993(68) & 1.19(11) \\
confined, $R=8$ & 0.0015(10) & 1.66(10) \\
stretched bulk, $R=3$ & 0.0031(06) &1.53(03) \\
stretched bulk, $R=8$ & 0.0051(06) & 1.46(17) \\
\hline
\end{tabular}
\end{table}

\begin{figure}
\centering
\includegraphics[width=0.48\textwidth]{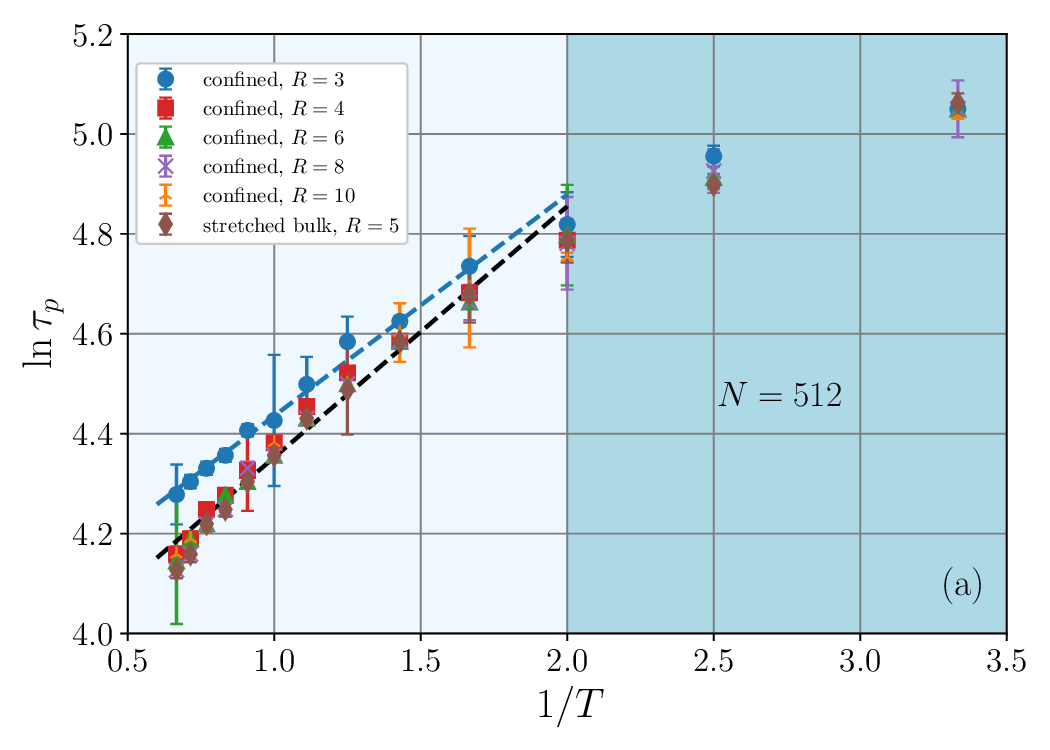}\\
\includegraphics[width=0.48\textwidth]{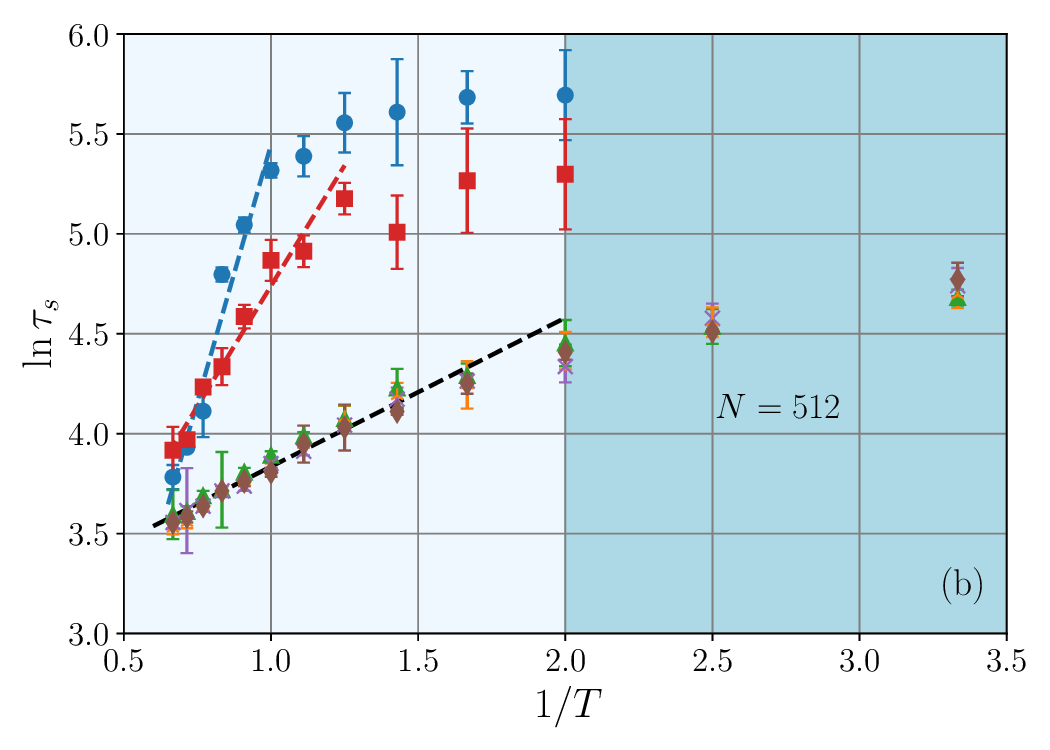}
\caption{(a) Arrhenius plots showing the temperature dependence of the pearl-necklace relaxation time $\tau_p$ of a collapsing polymer of chain length $N=512$ for different cases, as indicated. The dashed lines are the best fit straight lines, respectively, for $R=3$  and $R\ge 4$, representing the ansatz in Eq.\  \eqref{equation:arhenius}. (b) Arrhenius plots for the sausage-relaxation time $\tau_s$ for the same systems as in (a). The dashed lines again are fits for $\tau_s$, respectively, for the cases with $R=3$, $R=4$ and $R\ge6$, representing the ansatz in  Eq.\ \eqref{equation:arhenius}. The range of $T$ used for the fitting  are same as the range of the plotted dashed lines in respective cases.}
\label{fig:arrhenius_plot}
\end{figure}
\begin{table}[b!]
\centering
\caption{Results from fitting of the temperature dependence of $\tau_p$ using the ansatz in Eq. \eqref{equation:arhenius} for a collapsing polymer having $N=512$, under different conditions. }
\label{tab:arrhenius_table_tau_p}
\begin{tabular}{lll} 
\hline
\hline
~~~~~~Case & ~~~$\ln \tau_{\infty}$ &~~~$\frac{E_{\rm a}}{k_B}$ \\
\hline
confined, $R=3$ & 3.99(1) & 0.44(1) \\
confined, $R=4$ & 3.91(3) & 0.45(2) \\
confined, $R=6$ & 3.86(3) & 0.50(2) \\
confined, $R=8$ & 3.78(3) & 0.57(3) \\
confined, $R=10$ & 3.92(3) &0.45(3) \\
stretched bulk, $R=5$ & 3.78(3) & 0.56(2) \\
\hline
\end{tabular}
\end{table}

\par
To check the temperature dependence of the relaxation times, we perform simulations at different $T\in [0.3,1.5]$, for a polymer having $N=512$. In this regard, for the confined cases we present results for different values of $R$. For $R \le 4$ at low $T$ often the system gets trapped in metastable states making the extraction of $\tau_s$ extremely difficult. Hence, for these two cases we present results only for $T\ge 0.5$. For the stretched-bulk case we present results for $R=5$ only, as there the dynamics is largely independent of confinement. We quantify the temperature dependence through verification of Arrhenius behavior, which is a popular practice while studying the folding rates of proteins \cite{scalley1997,eaton2000fast,naganathan2007protein}. An Arhenius behavior characterizes the temperature dependence of a relaxation time as

\begin{equation}\label{tau_exp}
 \tau = \tau_{\infty}\exp\left(\frac{E_{\rm a}}{k_{\rm B} T}\right),
\end{equation}
where $E_{\rm a}$ is the related activation energy and $\tau_{\infty}$ is the amplitude of the $T$ dependence. On taking logarithm, the above relation transforms to

\begin{equation}
 \ln \tau = \ln \tau_{\infty} + \frac{E_{\rm a}}{k_{\rm B}} \cdot \frac{1}{T}.
 \label{equation:arhenius}
\end{equation}
Typically, one confirms an Arrhenius behavior if a plot of $\ln \tau$ shows a linear behavior as a function of the inverse temperature $1/T$. Such a behavior also allows one to have an estimate of $E_{\rm a}$ from the slope of the straight-line fit.
\begin{table}[t!]
\centering
\caption{Results from fitting of the temperature dependence of $\tau_s$ using the ansatz in Eq. \eqref{equation:arhenius} for a collapsing polymer having $N=512$, under different conditions.}
\label{tab:arrhenius_table_tau_s}
\begin{tabular}{lll} 
\hline
\hline
~~~~~~Case & ~~~~$\ln \tau_{\infty}$ &~~~~~$\frac{E_{\rm a}}{k_B}$ \\
\hline
confined, $R=3$ & 0.31(38) & 5.14(47) \\
confined, $R=4$ & 2.33(23) & 2.41(29) \\
confined, $R=6$ & 3.12(05) & 0.77(05) \\
confined, $R=8$ & 3.11(05) & 0.72(04) \\
confined, $R=10$ & 3.01(05) & 0.82(05) \\
stretched bulk, $R=5$ & 3.14(03) & 0.67(02) \\
\hline
\end{tabular}
\end{table}
\par
Figures\ \ref{fig:arrhenius_plot}(a) and (b), respectively, show the Arrhenius plots for $\tau_p$ and $\tau_s$ under different conditions, as mentioned. Except for the very small $T$ range, the data for both $\tau_p$ and $\tau_s$ in all the cases show a linear behavior confirming  Arrhenius behaviors. The deviation of the data for $T<0.5$ could possibly be due to the low-temperature driven early formation of sausage-like intermediates \cite{majumder2024temperature}. Hence, we fit Eq.\ \eqref{equation:arhenius} to the data of $\tau_p$ in the range marked by the lighter shade in Fig.\ \ref{fig:arrhenius_plot}(a). For $\tau_s$ the chosen fitting range is dependent on $R$. For $R\ge6$ we use the lightly shaded region in Fig.\ \ref{fig:arrhenius_plot}(b), and for the other cases the fitting ranges are same as the range in which the dashed lines are drawn. The results of the fitting exercises are tabulated in Tables\ \ref{tab:arrhenius_table_tau_p} and \ref{tab:arrhenius_table_tau_s}, respectively, for $\tau_p$ and $\tau_s$. The obtained $E_{\rm a}$ of the pearl-necklace stage for different cases have a very narrow range $E_{\rm a} \in [0.44,0.56]$, including the data for the stretched-bulk case, suggesting its independence on confinement. Comparison of  data in Tables\ \ref{tab:arrhenius_table_tau_p} and \ref{tab:arrhenius_table_tau_s} reveals that $E_{\rm a}$ for the sausage-relaxation stages are significantly larger than the pearl-necklace stage for smaller $R$. Unlike the pearl-necklace stage, the extracted $E_{\rm a}$ for the sausage relaxation in strongly ($R=3$ and $4$) confined cases are significantly (an order of magnitude) larger than that of a weakly ($R\ge6$) confined case. Along with that, it can also be noticed  that the stretched-bulk case has a slightly smaller $E_{\rm a}$ compared to the confined cases with large $R$. All these are implications of how the energy barriers of the sausage relaxation get modified  in the presence of confinement, leading to a $R$ dependent non-universal dynamics.

\begin{figure*}[t!]
\centering
\includegraphics[width=0.33\textwidth]{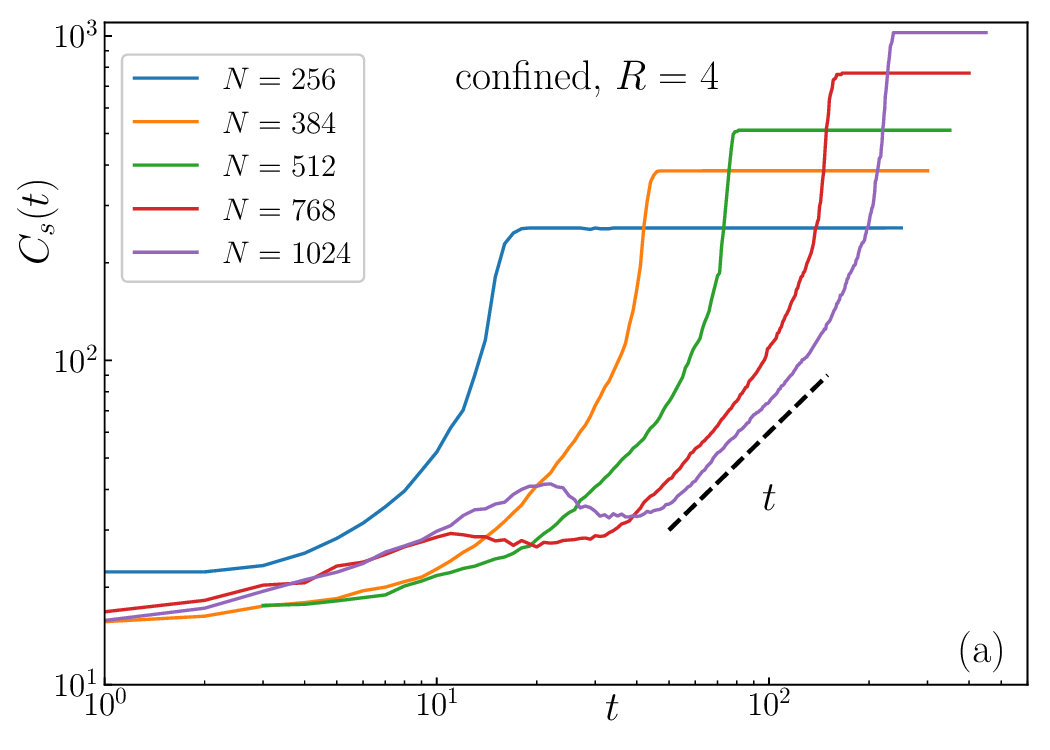}
\includegraphics[width=0.33\textwidth]{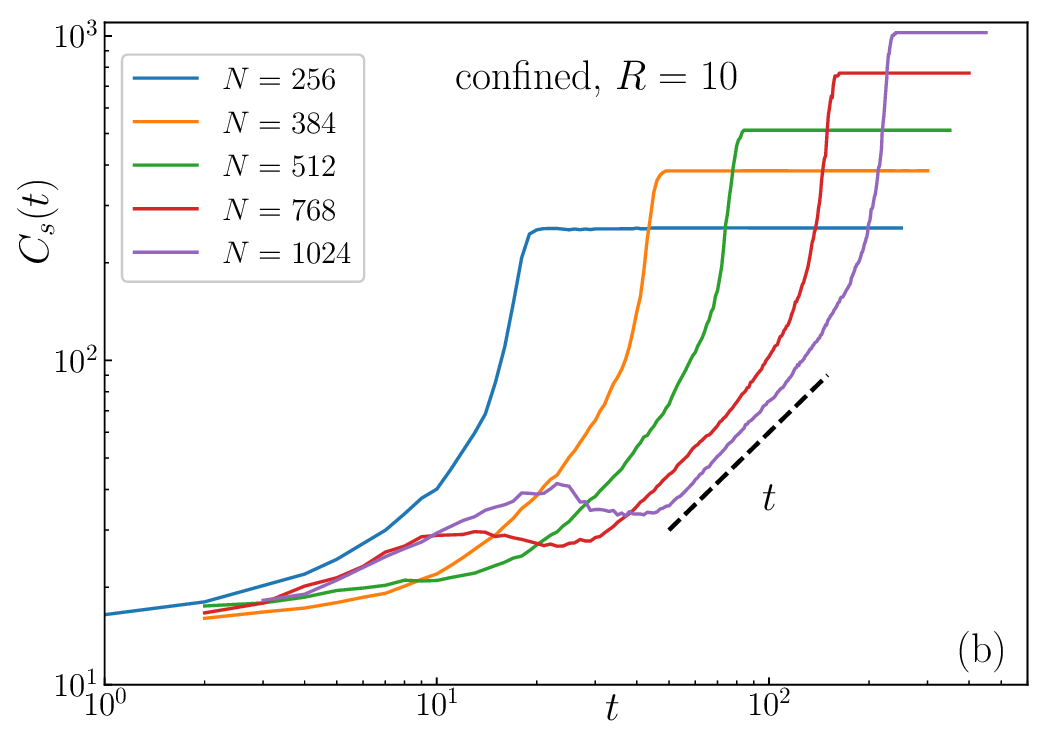}
\includegraphics[width=0.33\textwidth]{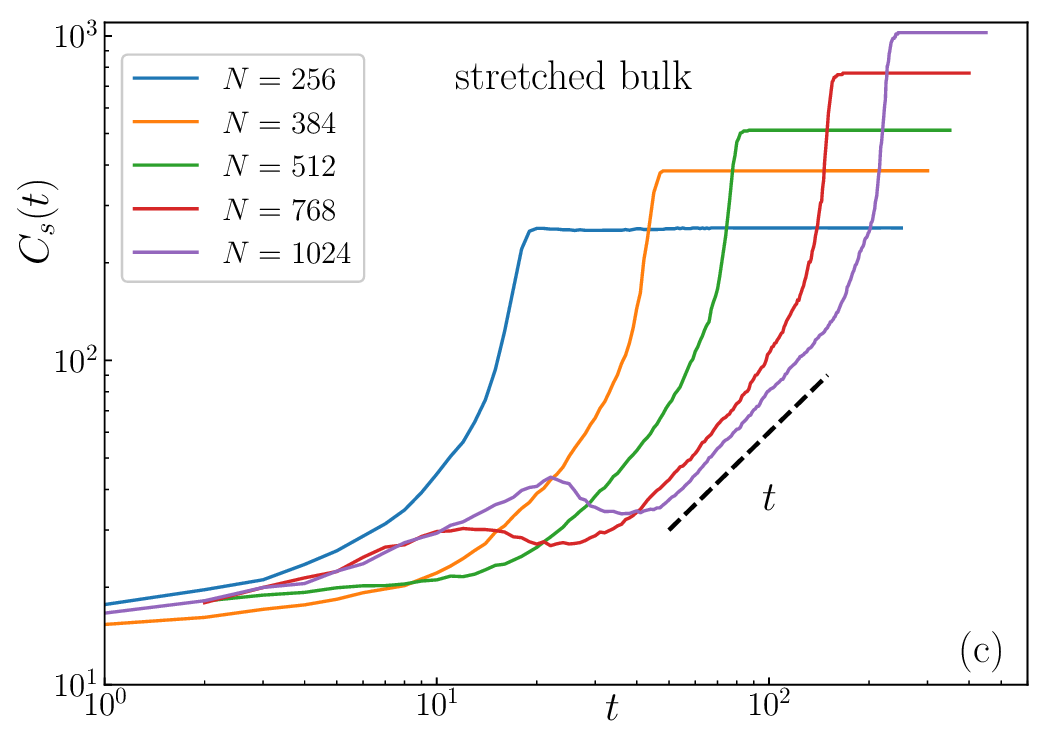}
\caption{Double-log scale plots of the time dependence of the average cluster size $C_s$ at $T=1.0$, for polymers collapsing in cylinders of radius (a) $R=4$, (b) $R=10$, and under (c) the stretched-bulk condition. In each case data for different chain lengths $N$ are presented. The dashed lines represent the linear scaling during the coarsening regime of the cluster growth.}\label{fig:cluster_size}
\end{figure*}

\subsection{Cluster-Growth Dynamics}
Now, we move onto the other important aspect of the collapse dynamics, i.e., the scaling of the time dependence of the cluster growth in the pearl-necklace stage. As already mentioned in the previous section, we identify each and every discrete clusters at a given time. The number of monomers within a cluster determines its size. From sizes of all the clusters present at a given time, we calculate the average cluster size as
\begin{equation}\label{cluster_growth}
 C_s(t) =\left \langle \frac{1}{n_c(t)} \sum_{k=1}^{n_c(t)} m_k \right \rangle.
\end{equation}
Figures \ref{fig:cluster_size}(a) and \ref{fig:cluster_size}(b) show the time dependence of $C_s(t)$, respectively, for polymers confined inside cylinders of $R=4$ and $10$. In Fig.\ \ref{fig:cluster_size}(c) we show the corresponding plots for the stretched-bulk case. 
In all the cases data for polymers with different $N\in[256,1024]$ at $T=1.0$ are presented. 
The behavior of $C_s(t)$ is almost same in all the cases. At early time there is a slow growth following which there is a linear increase before the data shoot up at late time. In the  initial slow-growth regime, nucleation or formation of discrete clusters occurs. The linear growth represents the coarsening regime, where the clusters merge with each other via diffusion along the axis of the cylinder. Once the clusters are considerably big, gradually the tension in the bridges of unclustered monomers becomes maximum. Eventually, the tension in the bridges pulls the clusters toward each other along the cylinder axis forming a single cluster. This gets manifested as a jump in the cluster growth curve at late time. At even later time, the saturation of the data around $C_s\approx N$ indicates that the pearl-necklace stage has ended and a single cluster is persisting containing all the monomers in it.
\begin{figure}[b!]
\centering
\includegraphics[width=0.48\textwidth]{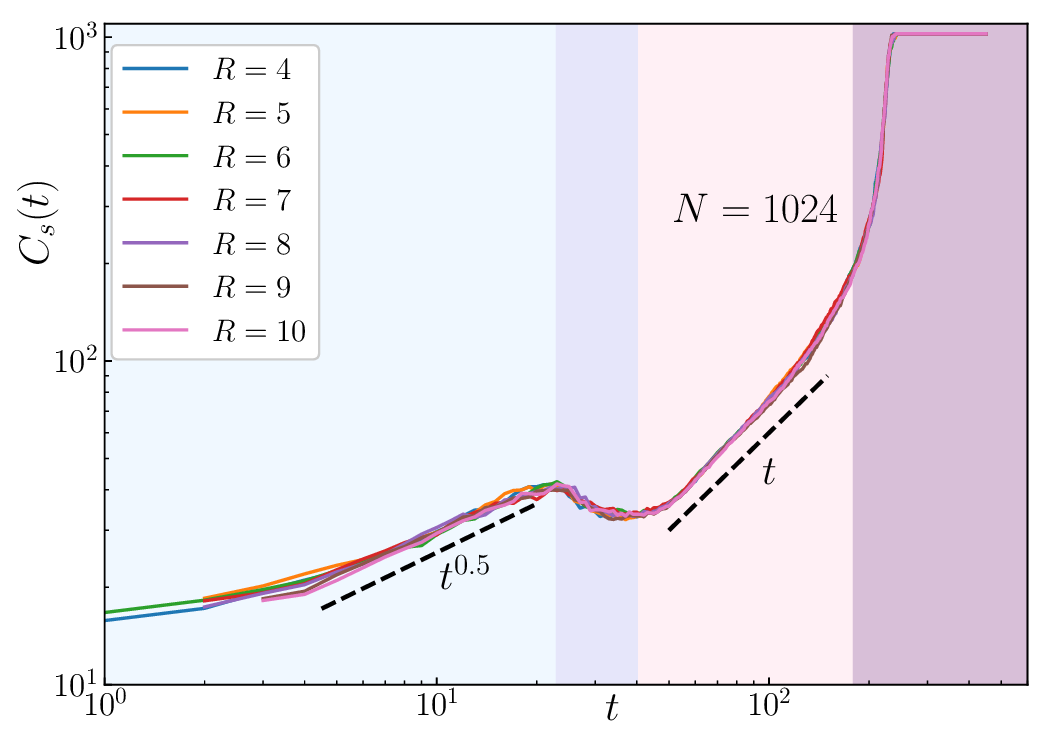}
\caption{Time-dependence of the average cluster size $C_s$ at $T=1.0$, for a polymer of chain length with $N=1024$, collapsing inside cylinders of different radius $R$. The dashed lines represent power-law scaling at different regimes, as indicated. The shades roughly correspond to various regimes of the cluster growth.}
\label{fig:cluster_size_r}
\end{figure}
\par
For examining the influence of confinement on the cluster-growth dynamics, in  Fig.\ \ref{fig:cluster_size_r} we present the time dependence of $C_s(t)$ for various $R$, at $T=1.0$, using a polymer with $N=1024$.  The overlap of data for all the cases suggests that despite variations in the strength of the cylindrical confinement, the growth dynamics is uniform. This implies that the cluster coalescence is mostly governed by local diffusive arrangements of the monomers,  which is rather weakly dependent on   the global geometric constraints. Along with the already discussed regimes, in Fig.\ \ref{fig:cluster_size_r} one also notices a hump in the data around $t\approx 20$. This can be  attributed to the fact that initially clusters are formed at the two ends of the polymer, which start growing by withdrawing monomers from the chain connecting them. This results in a slow-growth regime till $t\approx 20$, which roughly follows a $C_s(t)\sim t^{0.5}$ behavior. Afterwards, clusters start to develop uniformly throughout the chain, which are of smaller sizes, consequently reducing the average cluster size, i.e., $C_s$ for a while, and thereby giving a hump in the data. Eventually the system enters a proper coarsening regime, where the clusters grow linearly with time. Note that the hump in the data is also visible for $N\ge768$ in Fig.\ \ref{fig:cluster_size}. For shorter chains the above mentioned phenomenological explanation does not get manifested in the data, a signature of finite-size effects.
\begin{figure*}[t!]
\centering
\includegraphics[width=0.33\textwidth]{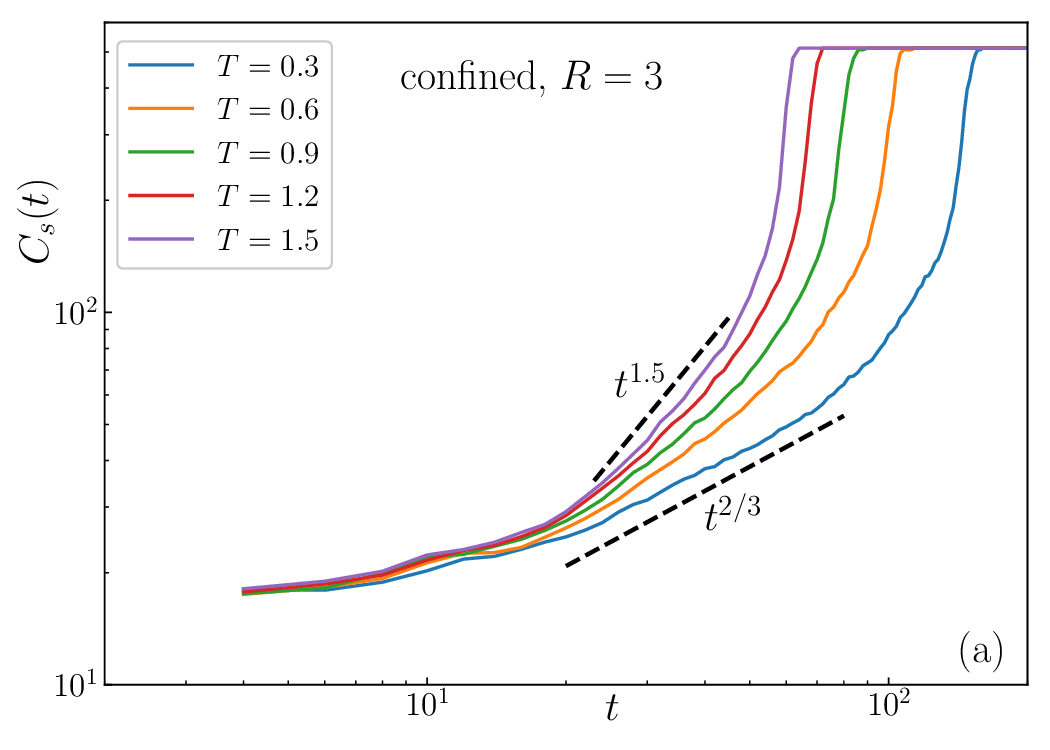} 
\includegraphics[width=0.33\textwidth]{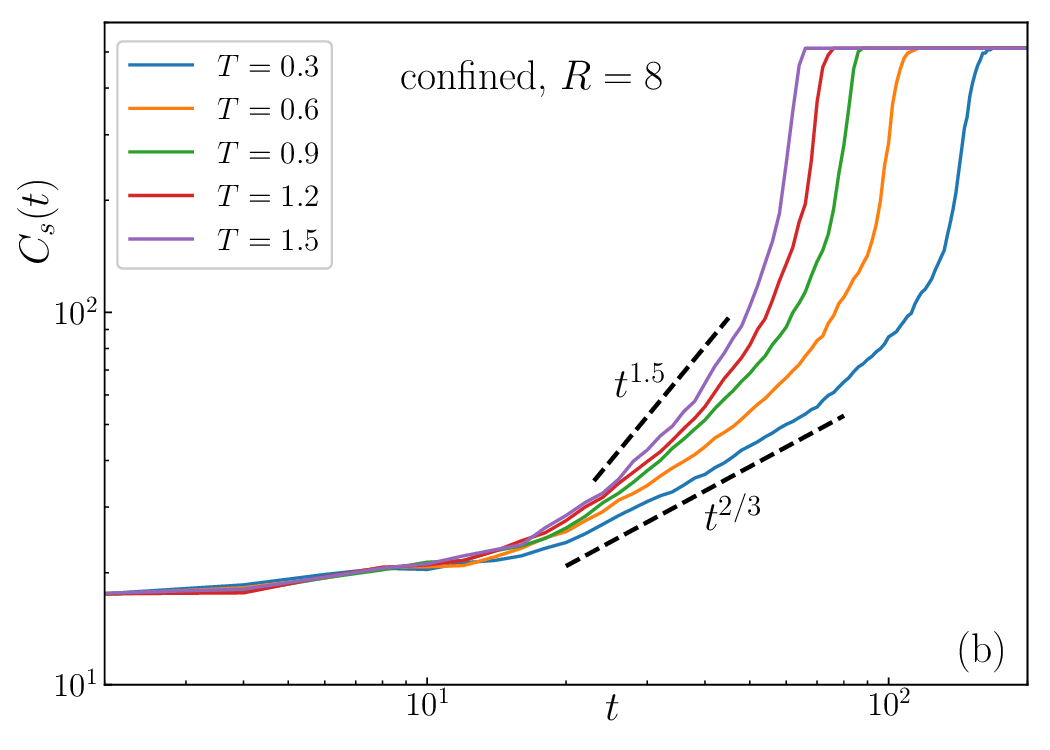} 
\includegraphics[width=0.33\textwidth]{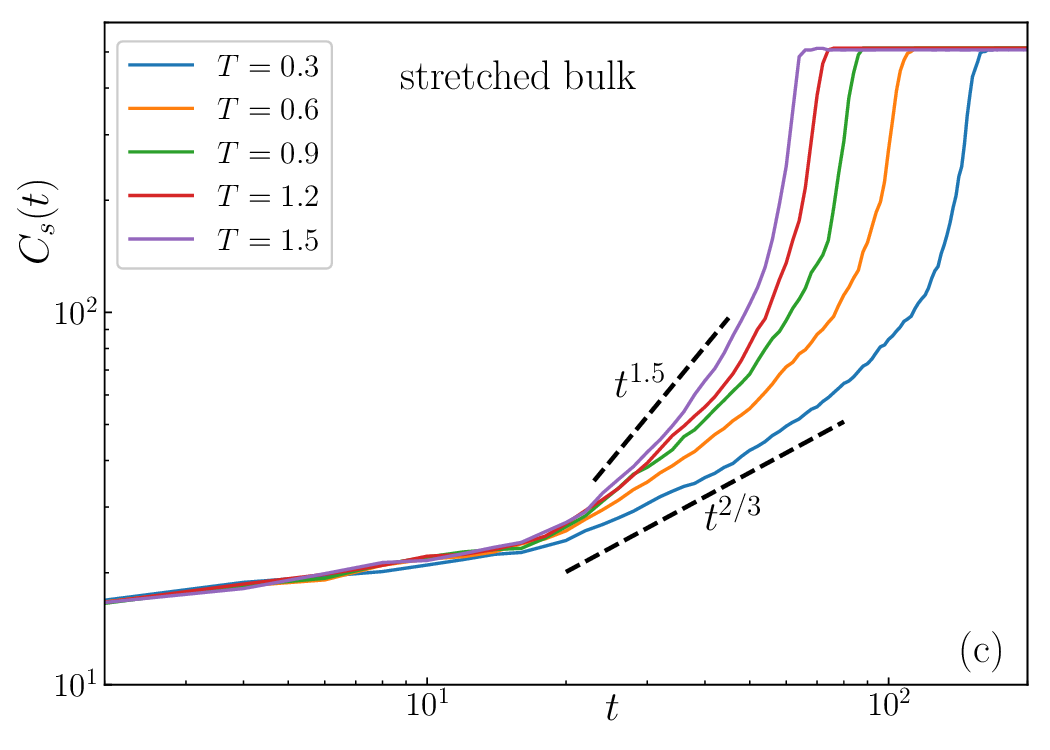}
\caption{Dependence of the scaling of the time dependence of cluster growth on temperature $T$, for a polymer collapsing inside a cylinder with radius (a) $R=3$ and (b) $8$. (c) Corresponding plots for the stretched-bulk case. The dashed lines represent different power-law behaviors in the coarsening regime, as indicated.}
\label{fig:cluster_size_temp}
\end{figure*}

To elucidate the effect of temperature on cluster-growth dynamics, in Figs.\ \ref{fig:cluster_size_temp}(a)-(c) we present corresponding plots of $C_s(t)$ at different $T$ for different cases. A clear dependence of cluster growth on $T$ is observed for all the cases. At early time, the data for all $T$ coincide with each other before they gradually start deviating, as the growth enters the coarsening regime. In this regime, the growth becomes faster as $T$ increases. At the lowest temperature ($T=0.3$), the data are consistent with a sub-linear power law behavior of $C_s(t) \sim t^{2/3}$. For higher $T$ the growth changes to a super-linear behavior, e.g., at the highest temperature ($T=1.5$) the data roughly follow a $C_s(t) \sim t^{1.5}$ behavior. For the usual-bulk case using the same polymer model, the reported growth exponent is $z\approx 1.67$ at $T=1.0$ \cite{majumder2025scale}. This suggests that growth dynamics is substantially slower even at high $T$ due to the influence of the confinement. The growth exponents for coarsening of spin or particle systems both in bulk \cite{majumder2017kinetics,majumder2013temperature,majumder2018universal} and in cylindrical confinement \cite{basu2016phase,basu2026}, typically, are independent of the quench temperature. This is in contradiction to the current observation. Plausibly, the complex combination of the intrinsic chain topology of the polymer  and the confinement is giving rise to this non-universal behavior with respect to $T$. We also notice an interesting fact from the growth dynamics of the stretched-bulk case, i.e. where the confinement is removed after equilibration. As there the polymer is free to collapse without any geometrical constraints, one expects it to exhibit  a cluster growth different from the confined cases. At all $T$, however, it exhibits a similar behavior. This implies that a structural memory of the initial stretched conformation is carried by the  polymer until a single cluster is formed. The same effect is certainly present in the confined cases as well. The memory effect is supposed to be $T$ dependent, and hence in all the cases it gets manifested in the growth dynamics too.

\section{Conclusion}\label{conclusion}

In this work we have investigated the collapse kinetics of homopolymers of different chain lengths $N$, confined inside cylinders of different radius $R$, using MD simulations. For comparison we have also studied the stretched-bulk case, where the confinement is removed after equilibration at high temperature. Comparison of these cases have allowed us to isolate the role of confinement geometry on the collapse kinetics.
\par
Unlike a homopolymer collapse in bulk, here, we have observed the presence of both the pearl-necklace and sausage-like intermediates highlighting, respectively, two distinct stages. 
The first stage is the pearl-necklace stage, which is characterized by the formation and growth of clusters of monomers, finally leading to a single sausage-like cluster. In the second stage, the sausage-like conformation tries to transform itself to a spherical globule. Since, the prevalent methods of monitoring the kinetics are not sufficient to capture the essence of the two distinct stages, we have relied on the asphericity of the individual clusters or pearls formed during the collapse to disentangle the related time scales. The pearl-necklace relaxation time $\tau_p$ is found to be independent of $R$ for a fixed $N$ at fixed temperature $T$. Also, the scaling $\tau_p \sim N^{z_p}$ has turned out to be quite robust with respect to $R$ having $z_p \approx 1.56$. Similarly, the activation energy $E_{\rm a}$ related to this stage is found to be varying within a very small range as the confinement changes from weak to strong. In contrast, the dynamics of the sausage-relaxation stage is firmly controlled by the strength of the confinement. For a fixed $N$, the sausage-relaxation time $\tau_s$ varies inversely with $R$ until it reaches a saturation for loose confinements. The power-law scaling $\tau_s \sim N^{z_s}$  is also dependent on the choice of $R$. While in a strong confinement $z_s\approx 1.19$, for a weak confinement $z_s\approx 1.55$. Similarly, $E_{\rm a}$ of this stage is found to be significantly higher in stronger confinements than in the weaker ones. 
\par
We have also studied the cluster-growth dynamics during the pearl-necklace stage. Surprisingly, the growth dynamics at a fixed $T$ shows universal behavior, irrespective of $R$. The cluster growth for large $N$ consists of multiple regimes. The initial slow-growth regime with $C_s(t) \sim t^{0.5}$ characterizes the growth of the clusters formed at the two ends.  Afterward, clusters start to develop uniformly along the chain. Following that there is a coarsening regime where the clusters grow almost linearly with time. Finally, in the last regime the growth accelerates due to the pulling of the clusters toward each other using the tension of the bridges of unclustered monomers. This gets manifested as a jump in the data. Our results from different $T$ reveal that although the presence of different growth regimes remain unaltered, the exponent of the power-law scaling of the cluster growth in the coarsening regime increases continuously with $T$.

\par
To build an experimental set up for the realization of the results obtained here, one can rely on glass or silicon made nanochannels having cylindrical cross-section of diameters ranging from $10$ nm to $100$ nm. In such a set up, conformations and dynamics of labeled DNA or synthetic polymers can be probed using time-resolved single-molecule spectroscopic techniques \cite{pollack2001time,tegenfeldt2004dynamics,reisner2005statics,reisner2007nanoconfinement,hofmann2012polymer,tress2013glassy}. In situ collapse of the polymer can be triggered via controlled changes in solvent quality, for example by introducing  multivalent counterions and depletion agents \cite{besteman2007role}. The temperature parameter in our simulations can be interpreted as a control to vary the solvent quality instead of a direct analog of the physical temperature in the above experimental scenarios. For that matter, a direct approach will be to use thermoresponsive polymers like PNIPAM which exhibits collapse transition under moderate temperature changes near the lower critical solution temperature \cite{adelsberger2012kinetics}.  All these proposed platforms present viable routes for a direct comparison of the simulation results obtained here with real experimental situations of polymer collapse inside cylindrical nanochannels.

\par
The cylindrical confinement considered here consists of a rigid repulsive wall. A real biological confinement is mostly soft in nature. For example, it is well known that the  cellular compartments are soft, and have ever changing shapes \cite{alberts1994molecular}. The dynamic change of cell shape in turn  alters the rate of various intra-cellular biophysical and biochemical processes involving biopolymers \cite{packer1963size,meyers2006potential,lizana2008controlling}. Inspired by that, as a next step we aim to investigate the collapse dynamics in a closed confinement with soft and dynamically deformable walls. We expect that the phenomenon observed here for the rigid wall, will get substantially modified in presence of soft wall, especially the sausage-relaxation stage. It would also be intriguing to look at the problem from a different perspective, i.e., how the conformational changes of the confined polymer trigger changes in shape of such a soft deformable compartment. We consider this also as one of our future endeavors.

\acknowledgments
The work was funded by the Anusandhan National Research Foundation (ANRF), Govt.\ of India through a Ramanujan Fellowship (File no.\ RJF/2021/000044). \\


\begin{thebibliography}{67}%
\makeatletter
\providecommand \@ifxundefined [1]{%
 \@ifx{#1\undefined}
}%
\providecommand \@ifnum [1]{%
 \ifnum #1\expandafter \@firstoftwo
 \else \expandafter \@secondoftwo
 \fi
}%
\providecommand \@ifx [1]{%
 \ifx #1\expandafter \@firstoftwo
 \else \expandafter \@secondoftwo
 \fi
}%
\providecommand \natexlab [1]{#1}%
\providecommand \enquote  [1]{``#1''}%
\providecommand \bibnamefont  [1]{#1}%
\providecommand \bibfnamefont [1]{#1}%
\providecommand \citenamefont [1]{#1}%
\providecommand \href@noop [0]{\@secondoftwo}%
\providecommand \href [0]{\begingroup \@sanitize@url \@href}%
\providecommand \@href[1]{\@@startlink{#1}\@@href}%
\providecommand \@@href[1]{\endgroup#1\@@endlink}%
\providecommand \@sanitize@url [0]{\catcode `\\12\catcode `\$12\catcode
  `\&12\catcode `\#12\catcode `\^12\catcode `\_12\catcode `\%12\relax}%
\providecommand \@@startlink[1]{}%
\providecommand \@@endlink[0]{}%
\providecommand \url  [0]{\begingroup\@sanitize@url \@url }%
\providecommand \@url [1]{\endgroup\@href {#1}{\urlprefix }}%
\providecommand \urlprefix  [0]{URL }%
\providecommand \Eprint [0]{\href }%
\providecommand \doibase [0]{http://dx.doi.org/}%
\providecommand \selectlanguage [0]{\@gobble}%
\providecommand \bibinfo  [0]{\@secondoftwo}%
\providecommand \bibfield  [0]{\@secondoftwo}%
\providecommand \translation [1]{[#1]}%
\providecommand \BibitemOpen [0]{}%
\providecommand \bibitemStop [0]{}%
\providecommand \bibitemNoStop [0]{.\EOS\space}%
\providecommand \EOS [0]{\spacefactor3000\relax}%
\providecommand \BibitemShut  [1]{\csname bibitem#1\endcsname}%
\let\auto@bib@innerbib\@empty
\bibitem [{\citenamefont {Flory}(1953)}]{flory1953principles}%
  \BibitemOpen
  \bibfield  {author} {\bibinfo {author} {\bibfnamefont {P.J.}\ \bibnamefont
  {Flory}},\ }\href@noop {} {\emph {\bibinfo {title} {Principles of Polymer
  Chemistry}}}\ (\bibinfo  {publisher} {Cornell university press},\ \bibinfo
  {year} {1953})\BibitemShut {NoStop}%
\bibitem [{\citenamefont {Rubenstein}\ and\ \citenamefont
  {Colby}(2003)}]{rubenstein2003}%
  \BibitemOpen
  \bibfield  {author} {\bibinfo {author} {\bibfnamefont {M.}~\bibnamefont
  {Rubenstein}}\ and\ \bibinfo {author} {\bibfnamefont {R.H.}\ \bibnamefont
  {Colby}},\ }\href@noop {} {\emph {\bibinfo {title} {Polymer Physics
  (Chemistry)}}}\ (\bibinfo  {publisher} {Oxford University Press, Oxford},\
  \bibinfo {year} {2003})\BibitemShut {NoStop}%
\bibitem [{\citenamefont {Camacho}\ and\ \citenamefont
  {Thirumalai}(1993)}]{camacho1993kinetics}%
  \BibitemOpen
  \bibfield  {author} {\bibinfo {author} {\bibfnamefont {C.J.}\ \bibnamefont
  {Camacho}}\ and\ \bibinfo {author} {\bibfnamefont {D.}~\bibnamefont
  {Thirumalai}},\ }\bibfield  {title} {\enquote {\bibinfo {title} {Kinetics and
  thermodynamics of folding in model proteins.}}\ }\href@noop {} {\bibfield
  {journal} {\bibinfo  {journal} {Proc. Natl. Acad. Sci. U.S.A.}\ }\textbf
  {\bibinfo {volume} {90}},\ \bibinfo {pages} {6369--6372} (\bibinfo {year}
  {1993})}\BibitemShut {NoStop}%
\bibitem [{\citenamefont {Pollack}\ \emph {et~al.}(2001)\citenamefont
  {Pollack}, \citenamefont {Tate}, \citenamefont {Finnefrock}, \citenamefont
  {Kalidas}, \citenamefont {Trotter}, \citenamefont {Darnton}, \citenamefont
  {Lurio}, \citenamefont {Austin}, \citenamefont {Batt}, \citenamefont
  {Gruner},\ and\ \citenamefont {Mochrie}}]{pollack2001time}%
  \BibitemOpen
  \bibfield  {author} {\bibinfo {author} {\bibfnamefont {L}~\bibnamefont
  {Pollack}}, \bibinfo {author} {\bibfnamefont {M.W.}\ \bibnamefont {Tate}},
  \bibinfo {author} {\bibfnamefont {A.C.}\ \bibnamefont {Finnefrock}}, \bibinfo
  {author} {\bibfnamefont {C.}~\bibnamefont {Kalidas}}, \bibinfo {author}
  {\bibfnamefont {S.}~\bibnamefont {Trotter}}, \bibinfo {author} {\bibfnamefont
  {N.C.}\ \bibnamefont {Darnton}}, \bibinfo {author} {\bibfnamefont
  {L.}~\bibnamefont {Lurio}}, \bibinfo {author} {\bibfnamefont {R.H.}\
  \bibnamefont {Austin}}, \bibinfo {author} {\bibfnamefont {C.A.}\ \bibnamefont
  {Batt}}, \bibinfo {author} {\bibfnamefont {S.M.}\ \bibnamefont {Gruner}}, \
  and\ \bibinfo {author} {\bibfnamefont {S.}~\bibnamefont {Mochrie}},\
  }\bibfield  {title} {\enquote {\bibinfo {title} {Time resolved collapse of a
  folding protein observed with small angle x-ray scattering},}\ }\href@noop {}
  {\bibfield  {journal} {\bibinfo  {journal} {Phys. Rev. Lett.}\ }\textbf
  {\bibinfo {volume} {86}},\ \bibinfo {pages} {4962} (\bibinfo {year}
  {2001})}\BibitemShut {NoStop}%
\bibitem [{\citenamefont {Sadqi}\ \emph {et~al.}(2003)\citenamefont {Sadqi},
  \citenamefont {Lapidus},\ and\ \citenamefont {Munoz}}]{sadqi2003fast}%
  \BibitemOpen
  \bibfield  {author} {\bibinfo {author} {\bibfnamefont {M.}~\bibnamefont
  {Sadqi}}, \bibinfo {author} {\bibfnamefont {L.J.}\ \bibnamefont {Lapidus}}, \
  and\ \bibinfo {author} {\bibfnamefont {V.}~\bibnamefont {Munoz}},\ }\bibfield
   {title} {\enquote {\bibinfo {title} {How fast is protein hydrophobic
  collapse?}}\ }\href@noop {} {\bibfield  {journal} {\bibinfo  {journal} {Proc.
  Natl. Acad. Sci. U.S.A.}\ }\textbf {\bibinfo {volume} {100}},\ \bibinfo
  {pages} {12117--12122} (\bibinfo {year} {2003})}\BibitemShut {NoStop}%
\bibitem [{\citenamefont {Haran}(2012)}]{haran2012}%
  \BibitemOpen
  \bibfield  {author} {\bibinfo {author} {\bibfnamefont {G.}~\bibnamefont
  {Haran}},\ }\bibfield  {title} {\enquote {\bibinfo {title} {How, when and why
  proteins collapse: {T}he relation to folding},}\ }\href@noop {} {\bibfield
  {journal} {\bibinfo  {journal} {Curr. Opin. Struct. Biol.}\ }\textbf
  {\bibinfo {volume} {22}},\ \bibinfo {pages} {14--20} (\bibinfo {year}
  {2012})}\BibitemShut {NoStop}%
\bibitem [{\citenamefont {Udgaonkar}(2013)}]{udgaonkar2013}%
  \BibitemOpen
  \bibfield  {author} {\bibinfo {author} {\bibfnamefont {J.B.}\ \bibnamefont
  {Udgaonkar}},\ }\bibfield  {title} {\enquote {\bibinfo {title} {Polypeptide
  chain collapse and protein folding},}\ }\href@noop {} {\bibfield  {journal}
  {\bibinfo  {journal} {Arch. Biochem. Biophys.}\ }\textbf {\bibinfo {volume}
  {531}},\ \bibinfo {pages} {24--33} (\bibinfo {year} {2013})}\BibitemShut
  {NoStop}%
\bibitem [{\citenamefont {Reddy}\ and\ \citenamefont
  {Thirumalai}(2017)}]{reddy2017collapse}%
  \BibitemOpen
  \bibfield  {author} {\bibinfo {author} {\bibfnamefont {G.}~\bibnamefont
  {Reddy}}\ and\ \bibinfo {author} {\bibfnamefont {D.}~\bibnamefont
  {Thirumalai}},\ }\bibfield  {title} {\enquote {\bibinfo {title} {Collapse
  precedes folding in denaturant-dependent assembly of ubiquitin},}\
  }\href@noop {} {\bibfield  {journal} {\bibinfo  {journal} {J. Phys. Chem. B}\
  }\textbf {\bibinfo {volume} {121}},\ \bibinfo {pages} {995--1009} (\bibinfo
  {year} {2017})}\BibitemShut {NoStop}%
\bibitem [{\citenamefont {de~Gennes}(1985)}]{de1985kinetics}%
  \BibitemOpen
  \bibfield  {author} {\bibinfo {author} {\bibfnamefont {P.-G}\ \bibnamefont
  {de~Gennes}},\ }\bibfield  {title} {\enquote {\bibinfo {title} {Kinetics of
  collapse for a flexible coil},}\ }\href@noop {} {\bibfield  {journal}
  {\bibinfo  {journal} {J. Phys. Lett.}\ }\textbf {\bibinfo {volume} {46}},\
  \bibinfo {pages} {639--642} (\bibinfo {year} {1985})}\BibitemShut {NoStop}%
\bibitem [{\citenamefont {Byrne}\ \emph {et~al.}(1995)\citenamefont {Byrne},
  \citenamefont {Kiernan}, \citenamefont {Green},\ and\ \citenamefont
  {Dawson}}]{byrne1995kinetics}%
  \BibitemOpen
  \bibfield  {author} {\bibinfo {author} {\bibfnamefont {A.}~\bibnamefont
  {Byrne}}, \bibinfo {author} {\bibfnamefont {P.}~\bibnamefont {Kiernan}},
  \bibinfo {author} {\bibfnamefont {D.}~\bibnamefont {Green}}, \ and\ \bibinfo
  {author} {\bibfnamefont {K.A.}\ \bibnamefont {Dawson}},\ }\bibfield  {title}
  {\enquote {\bibinfo {title} {Kinetics of homopolymer collapse},}\ }\href@noop
  {} {\bibfield  {journal} {\bibinfo  {journal} {J. Chem. Phys.}\ }\textbf
  {\bibinfo {volume} {102}},\ \bibinfo {pages} {573--577} (\bibinfo {year}
  {1995})}\BibitemShut {NoStop}%
\bibitem [{\citenamefont {Kuznetsov}\ \emph {et~al.}(1996)\citenamefont
  {Kuznetsov}, \citenamefont {Timoshenko},\ and\ \citenamefont
  {Dawson}}]{kuznetsov1996kinetic}%
  \BibitemOpen
  \bibfield  {author} {\bibinfo {author} {\bibfnamefont {Y.A.}\ \bibnamefont
  {Kuznetsov}}, \bibinfo {author} {\bibfnamefont {E.G.}\ \bibnamefont
  {Timoshenko}}, \ and\ \bibinfo {author} {\bibfnamefont {K.A.}\ \bibnamefont
  {Dawson}},\ }\bibfield  {title} {\enquote {\bibinfo {title} {Kinetic laws at
  the collapse transition of a homopolymer},}\ }\href@noop {} {\bibfield
  {journal} {\bibinfo  {journal} {J. Chem. Phys.}\ }\textbf {\bibinfo {volume}
  {104}},\ \bibinfo {pages} {3338--3347} (\bibinfo {year} {1996})}\BibitemShut
  {NoStop}%
\bibitem [{\citenamefont {Halperin}\ and\ \citenamefont
  {Goldbart}(2000)}]{halperin2000early}%
  \BibitemOpen
  \bibfield  {author} {\bibinfo {author} {\bibfnamefont {A.}~\bibnamefont
  {Halperin}}\ and\ \bibinfo {author} {\bibfnamefont {P.M.}\ \bibnamefont
  {Goldbart}},\ }\bibfield  {title} {\enquote {\bibinfo {title} {Early stages
  of homopolymer collapse},}\ }\href@noop {} {\bibfield  {journal} {\bibinfo
  {journal} {Phys. Rev. E}\ }\textbf {\bibinfo {volume} {61}},\ \bibinfo
  {pages} {565} (\bibinfo {year} {2000})}\BibitemShut {NoStop}%
\bibitem [{\citenamefont {Pitard}\ and\ \citenamefont
  {Bouchaud}(2001)}]{pitard2001glassy}%
  \BibitemOpen
  \bibfield  {author} {\bibinfo {author} {\bibfnamefont {E.}~\bibnamefont
  {Pitard}}\ and\ \bibinfo {author} {\bibfnamefont {J.-P.}\ \bibnamefont
  {Bouchaud}},\ }\bibfield  {title} {\enquote {\bibinfo {title} {Glassy effects
  in the swelling/collapse dynamics of homogeneous polymers},}\ }\href@noop {}
  {\bibfield  {journal} {\bibinfo  {journal} {Euro. Phys. J. E}\ }\textbf
  {\bibinfo {volume} {5}},\ \bibinfo {pages} {133--148} (\bibinfo {year}
  {2001})}\BibitemShut {NoStop}%
\bibitem [{\citenamefont {Dokholyan}\ \emph {et~al.}(2002)\citenamefont
  {Dokholyan}, \citenamefont {Pitard}, \citenamefont {Buldyrev},\ and\
  \citenamefont {Stanley}}]{dokholyan2002}%
  \BibitemOpen
  \bibfield  {author} {\bibinfo {author} {\bibfnamefont {N.V.}\ \bibnamefont
  {Dokholyan}}, \bibinfo {author} {\bibfnamefont {E.}~\bibnamefont {Pitard}},
  \bibinfo {author} {\bibfnamefont {S.V.}\ \bibnamefont {Buldyrev}}, \ and\
  \bibinfo {author} {\bibfnamefont {H.E.}\ \bibnamefont {Stanley}},\ }\bibfield
   {title} {\enquote {\bibinfo {title} {Glassy behavior of a homopolymer from
  molecular dynamics simulations},}\ }\href@noop {} {\bibfield  {journal}
  {\bibinfo  {journal} {Phys. Rev. E}\ }\textbf {\bibinfo {volume} {65}},\
  \bibinfo {pages} {030801} (\bibinfo {year} {2002})}\BibitemShut {NoStop}%
\bibitem [{\citenamefont {Abrams}\ \emph {et~al.}(2002)\citenamefont {Abrams},
  \citenamefont {Lee},\ and\ \citenamefont {Obukhov}}]{abrams2002collapse}%
  \BibitemOpen
  \bibfield  {author} {\bibinfo {author} {\bibfnamefont {C.F.}\ \bibnamefont
  {Abrams}}, \bibinfo {author} {\bibfnamefont {N.-K}\ \bibnamefont {Lee}}, \
  and\ \bibinfo {author} {\bibfnamefont {S.P.}\ \bibnamefont {Obukhov}},\
  }\bibfield  {title} {\enquote {\bibinfo {title} {Collapse dynamics of a
  polymer chain: Theory and simulation},}\ }\href@noop {} {\bibfield  {journal}
  {\bibinfo  {journal} {Europhys. Lett.}\ }\textbf {\bibinfo {volume} {59}},\
  \bibinfo {pages} {391} (\bibinfo {year} {2002})}\BibitemShut {NoStop}%
\bibitem [{\citenamefont {Kikuchi}\ \emph {et~al.}(2005)\citenamefont
  {Kikuchi}, \citenamefont {Ryder}, \citenamefont {Pooley},\ and\ \citenamefont
  {Yeomans}}]{kikuchi2005kinetics}%
  \BibitemOpen
  \bibfield  {author} {\bibinfo {author} {\bibfnamefont {N.}~\bibnamefont
  {Kikuchi}}, \bibinfo {author} {\bibfnamefont {J.F.}\ \bibnamefont {Ryder}},
  \bibinfo {author} {\bibfnamefont {C.M.}\ \bibnamefont {Pooley}}, \ and\
  \bibinfo {author} {\bibfnamefont {J.M.}\ \bibnamefont {Yeomans}},\ }\bibfield
   {title} {\enquote {\bibinfo {title} {Kinetics of the polymer collapse
  transition: The role of hydrodynamics},}\ }\href@noop {} {\bibfield
  {journal} {\bibinfo  {journal} {Phys. Rev. E}\ }\textbf {\bibinfo {volume}
  {71}},\ \bibinfo {pages} {061804} (\bibinfo {year} {2005})}\BibitemShut
  {NoStop}%
\bibitem [{\citenamefont {Xu}\ \emph {et~al.}(2006)\citenamefont {Xu},
  \citenamefont {Zhu}, \citenamefont {Luo}, \citenamefont {Wu},\ and\
  \citenamefont {Liu}}]{xu2006first}%
  \BibitemOpen
  \bibfield  {author} {\bibinfo {author} {\bibfnamefont {J.}~\bibnamefont
  {Xu}}, \bibinfo {author} {\bibfnamefont {Z.}~\bibnamefont {Zhu}}, \bibinfo
  {author} {\bibfnamefont {S.}~\bibnamefont {Luo}}, \bibinfo {author}
  {\bibfnamefont {C.}~\bibnamefont {Wu}}, \ and\ \bibinfo {author}
  {\bibfnamefont {S.}~\bibnamefont {Liu}},\ }\bibfield  {title} {\enquote
  {\bibinfo {title} {First observation of two-stage collapsing kinetics of a
  single synthetic polymer chain},}\ }\href@noop {} {\bibfield  {journal}
  {\bibinfo  {journal} {Phys. Rev. Lett.}\ }\textbf {\bibinfo {volume} {96}},\
  \bibinfo {pages} {027802} (\bibinfo {year} {2006})}\BibitemShut {NoStop}%
\bibitem [{\citenamefont {Ye}\ \emph {et~al.}(2007)\citenamefont {Ye},
  \citenamefont {Lu}, \citenamefont {Shen}, \citenamefont {Ding}, \citenamefont
  {Liu}, \citenamefont {Zhang},\ and\ \citenamefont {Wu}}]{ye2007many}%
  \BibitemOpen
  \bibfield  {author} {\bibinfo {author} {\bibfnamefont {X.}~\bibnamefont
  {Ye}}, \bibinfo {author} {\bibfnamefont {Y.}~\bibnamefont {Lu}}, \bibinfo
  {author} {\bibfnamefont {L.}~\bibnamefont {Shen}}, \bibinfo {author}
  {\bibfnamefont {Y.}~\bibnamefont {Ding}}, \bibinfo {author} {\bibfnamefont
  {S.}~\bibnamefont {Liu}}, \bibinfo {author} {\bibfnamefont {G.}~\bibnamefont
  {Zhang}}, \ and\ \bibinfo {author} {\bibfnamefont {C.}~\bibnamefont {Wu}},\
  }\bibfield  {title} {\enquote {\bibinfo {title} {How many stages in the
  coil-to-globule transition of linear homopolymer chains in a dilute
  solution?}}\ }\href@noop {} {\bibfield  {journal} {\bibinfo  {journal}
  {Macromolecules}\ }\textbf {\bibinfo {volume} {40}},\ \bibinfo {pages}
  {4750--4752} (\bibinfo {year} {2007})}\BibitemShut {NoStop}%
\bibitem [{\citenamefont {Majumder}\ and\ \citenamefont
  {Janke}(2015)}]{majumder2015cluster}%
  \BibitemOpen
  \bibfield  {author} {\bibinfo {author} {\bibfnamefont {S.}~\bibnamefont
  {Majumder}}\ and\ \bibinfo {author} {\bibfnamefont {W.}~\bibnamefont
  {Janke}},\ }\bibfield  {title} {\enquote {\bibinfo {title} {Cluster
  coarsening during polymer collapse: Finite-size scaling analysis},}\
  }\href@noop {} {\bibfield  {journal} {\bibinfo  {journal} {Europhys. Lett.}\
  }\textbf {\bibinfo {volume} {110}},\ \bibinfo {pages} {58001} (\bibinfo
  {year} {2015})}\BibitemShut {NoStop}%
\bibitem [{\citenamefont {Majumder}\ and\ \citenamefont
  {Janke}(2016{\natexlab{a}})}]{majumder2016evidence}%
  \BibitemOpen
  \bibfield  {author} {\bibinfo {author} {\bibfnamefont {S.}~\bibnamefont
  {Majumder}}\ and\ \bibinfo {author} {\bibfnamefont {W.}~\bibnamefont
  {Janke}},\ }\bibfield  {title} {\enquote {\bibinfo {title} {Evidence of aging
  and dynamic scaling in the collapse of a polymer},}\ }\href@noop {}
  {\bibfield  {journal} {\bibinfo  {journal} {Phys. Rev. E}\ }\textbf {\bibinfo
  {volume} {93}},\ \bibinfo {pages} {032506} (\bibinfo {year}
  {2016}{\natexlab{a}})}\BibitemShut {NoStop}%
\bibitem [{\citenamefont {Majumder}\ \emph {et~al.}(2017)\citenamefont
  {Majumder}, \citenamefont {Zierenberg},\ and\ \citenamefont
  {Janke}}]{majumder2017kinetics}%
  \BibitemOpen
  \bibfield  {author} {\bibinfo {author} {\bibfnamefont {S.}~\bibnamefont
  {Majumder}}, \bibinfo {author} {\bibfnamefont {J.}~\bibnamefont
  {Zierenberg}}, \ and\ \bibinfo {author} {\bibfnamefont {W.}~\bibnamefont
  {Janke}},\ }\bibfield  {title} {\enquote {\bibinfo {title} {Kinetics of
  polymer collapse: Effect of temperature on cluster growth and aging},}\
  }\href@noop {} {\bibfield  {journal} {\bibinfo  {journal} {Soft Matter}\
  }\textbf {\bibinfo {volume} {13}},\ \bibinfo {pages} {1276--1290} (\bibinfo
  {year} {2017})}\BibitemShut {NoStop}%
\bibitem [{\citenamefont {Christiansen}\ \emph {et~al.}(2017)\citenamefont
  {Christiansen}, \citenamefont {Majumder},\ and\ \citenamefont
  {Janke}}]{christiansen2017coarsening}%
  \BibitemOpen
  \bibfield  {author} {\bibinfo {author} {\bibfnamefont {H.}~\bibnamefont
  {Christiansen}}, \bibinfo {author} {\bibfnamefont {S.}~\bibnamefont
  {Majumder}}, \ and\ \bibinfo {author} {\bibfnamefont {W.}~\bibnamefont
  {Janke}},\ }\bibfield  {title} {\enquote {\bibinfo {title} {Coarsening and
  aging of lattice polymers: Influence of bond fluctuations},}\ }\href@noop {}
  {\bibfield  {journal} {\bibinfo  {journal} {J. Chem. Phys.}\ }\textbf
  {\bibinfo {volume} {147}} (\bibinfo {year} {2017})}\BibitemShut {NoStop}%
\bibitem [{\citenamefont {Majumder}\ \emph {et~al.}(2019)\citenamefont
  {Majumder}, \citenamefont {Hansmann},\ and\ \citenamefont
  {Janke}}]{majumder2019pearl}%
  \BibitemOpen
  \bibfield  {author} {\bibinfo {author} {\bibfnamefont {S.}~\bibnamefont
  {Majumder}}, \bibinfo {author} {\bibfnamefont {U.H.E.}\ \bibnamefont
  {Hansmann}}, \ and\ \bibinfo {author} {\bibfnamefont {W.}~\bibnamefont
  {Janke}},\ }\bibfield  {title} {\enquote {\bibinfo {title}
  {Pearl-necklace-like local ordering drives polypeptide collapse},}\
  }\href@noop {} {\bibfield  {journal} {\bibinfo  {journal} {Macromolecules}\
  }\textbf {\bibinfo {volume} {52}},\ \bibinfo {pages} {5491--5498} (\bibinfo
  {year} {2019})}\BibitemShut {NoStop}%
\bibitem [{\citenamefont {Majumder}\ \emph {et~al.}(2020)\citenamefont
  {Majumder}, \citenamefont {Christiansen},\ and\ \citenamefont
  {Janke}}]{majumder2020understanding}%
  \BibitemOpen
  \bibfield  {author} {\bibinfo {author} {\bibfnamefont {S.}~\bibnamefont
  {Majumder}}, \bibinfo {author} {\bibfnamefont {H.}~\bibnamefont
  {Christiansen}}, \ and\ \bibinfo {author} {\bibfnamefont {W.}~\bibnamefont
  {Janke}},\ }\bibfield  {title} {\enquote {\bibinfo {title} {Understanding
  nonequilibrium scaling laws governing collapse of a polymer},}\ }\href@noop
  {} {\bibfield  {journal} {\bibinfo  {journal} {Eur. Phys. J. B}\ }\textbf
  {\bibinfo {volume} {93}},\ \bibinfo {pages} {142} (\bibinfo {year}
  {2020})}\BibitemShut {NoStop}%
\bibitem [{\citenamefont {Schneider}\ \emph {et~al.}(2020)\citenamefont
  {Schneider}, \citenamefont {Meinel}, \citenamefont {Dittmar},\ and\
  \citenamefont {M\"uller-Plathe}}]{schneider2020}%
  \BibitemOpen
  \bibfield  {author} {\bibinfo {author} {\bibfnamefont {J.}~\bibnamefont
  {Schneider}}, \bibinfo {author} {\bibfnamefont {M.K.}\ \bibnamefont
  {Meinel}}, \bibinfo {author} {\bibfnamefont {H.}~\bibnamefont {Dittmar}}, \
  and\ \bibinfo {author} {\bibfnamefont {F.}~\bibnamefont {M\"uller-Plathe}},\
  }\bibfield  {title} {\enquote {\bibinfo {title} {Different stages of
  polymer-chain collapse following solvent quenching--scaling relations from
  dissipative particle dynamics simulations},}\ }\href@noop {} {\bibfield
  {journal} {\bibinfo  {journal} {Macromolecules}\ }\textbf {\bibinfo {volume}
  {53}},\ \bibinfo {pages} {8889--8900} (\bibinfo {year} {2020})}\BibitemShut
  {NoStop}%
\bibitem [{\citenamefont {Majumder}\ \emph {et~al.}(2024)\citenamefont
  {Majumder}, \citenamefont {Christiansen},\ and\ \citenamefont
  {Janke}}]{majumder2024temperature}%
  \BibitemOpen
  \bibfield  {author} {\bibinfo {author} {\bibfnamefont {S.}~\bibnamefont
  {Majumder}}, \bibinfo {author} {\bibfnamefont {H.}~\bibnamefont
  {Christiansen}}, \ and\ \bibinfo {author} {\bibfnamefont {W.}~\bibnamefont
  {Janke}},\ }\bibfield  {title} {\enquote {\bibinfo {title} {Temperature and
  viscosity tune the intermediates during the collapse of a polymer},}\
  }\href@noop {} {\bibfield  {journal} {\bibinfo  {journal} {Macromolecules}\
  }\textbf {\bibinfo {volume} {57}},\ \bibinfo {pages} {10586--10599} (\bibinfo
  {year} {2024})}\BibitemShut {NoStop}%
\bibitem [{\citenamefont {Majumder}\ and\ \citenamefont
  {Chakraborty}(2025)}]{majumder2025scale}%
  \BibitemOpen
  \bibfield  {author} {\bibinfo {author} {\bibfnamefont {S.}~\bibnamefont
  {Majumder}}\ and\ \bibinfo {author} {\bibfnamefont {S.}~\bibnamefont
  {Chakraborty}},\ }\bibfield  {title} {\enquote {\bibinfo {title} {Scale-free
  cluster--cluster aggregation during polymer collapse},}\ }\href@noop {}
  {\bibfield  {journal} {\bibinfo  {journal} {Macromolecules}\ }\textbf
  {\bibinfo {volume} {58}},\ \bibinfo {pages} {10212--10223} (\bibinfo {year}
  {2025})}\BibitemShut {NoStop}%
\bibitem [{\citenamefont {De~Gennes}(1979)}]{de1979scaling}%
  \BibitemOpen
  \bibfield  {author} {\bibinfo {author} {\bibfnamefont {P.G.}\ \bibnamefont
  {De~Gennes}},\ }\href@noop {} {\emph {\bibinfo {title} {Scaling Concepts in
  Polymer Physics}}}\ (\bibinfo  {publisher} {Cornell university press},\
  \bibinfo {year} {1979})\BibitemShut {NoStop}%
\bibitem [{\citenamefont {Milchev}(2011)}]{milchev2011}%
  \BibitemOpen
  \bibfield  {author} {\bibinfo {author} {\bibfnamefont {A.}~\bibnamefont
  {Milchev}},\ }\bibfield  {title} {\enquote {\bibinfo {title} {Single-polymer
  dynamics under constraints: scaling theory and computer experiment},}\
  }\href@noop {} {\bibfield  {journal} {\bibinfo  {journal} {J. Phys.: Condens.
  Matt.}\ }\textbf {\bibinfo {volume} {23}},\ \bibinfo {pages} {103101}
  (\bibinfo {year} {2011})}\BibitemShut {NoStop}%
\bibitem [{\citenamefont {Reisner}\ \emph {et~al.}(2012)\citenamefont
  {Reisner}, \citenamefont {Pedersen},\ and\ \citenamefont
  {Austin}}]{reisner2012dna}%
  \BibitemOpen
  \bibfield  {author} {\bibinfo {author} {\bibfnamefont {W.}~\bibnamefont
  {Reisner}}, \bibinfo {author} {\bibfnamefont {J.N.}\ \bibnamefont
  {Pedersen}}, \ and\ \bibinfo {author} {\bibfnamefont {R.H.}\ \bibnamefont
  {Austin}},\ }\bibfield  {title} {\enquote {\bibinfo {title} {{DNA}
  confinement in nanochannels: physics and biological applications},}\
  }\href@noop {} {\bibfield  {journal} {\bibinfo  {journal} {Rep. Prog. Phys.}\
  }\textbf {\bibinfo {volume} {75}},\ \bibinfo {pages} {106601} (\bibinfo
  {year} {2012})}\BibitemShut {NoStop}%
\bibitem [{\citenamefont {Wang}(2017)}]{wang201750}%
  \BibitemOpen
  \bibfield  {author} {\bibinfo {author} {\bibfnamefont {Zhen-Gang}\
  \bibnamefont {Wang}},\ }\bibfield  {title} {\enquote {\bibinfo {title} {{50th
  Anniversary Perspective: Polymer Conformation -- A Pedagogical Review}},}\
  }\href@noop {} {\bibfield  {journal} {\bibinfo  {journal} {Macromolecules}\
  }\textbf {\bibinfo {volume} {50}},\ \bibinfo {pages} {9073--9114} (\bibinfo
  {year} {2017})}\BibitemShut {NoStop}%
\bibitem [{\citenamefont {Perkins}\ \emph {et~al.}(1997)\citenamefont
  {Perkins}, \citenamefont {Smith},\ and\ \citenamefont
  {Chu}}]{perkins1997single}%
  \BibitemOpen
  \bibfield  {author} {\bibinfo {author} {\bibfnamefont {T.T.}\ \bibnamefont
  {Perkins}}, \bibinfo {author} {\bibfnamefont {D.E.}\ \bibnamefont {Smith}}, \
  and\ \bibinfo {author} {\bibfnamefont {S.}~\bibnamefont {Chu}},\ }\bibfield
  {title} {\enquote {\bibinfo {title} {Single polymer dynamics in an
  elongational flow},}\ }\href@noop {} {\bibfield  {journal} {\bibinfo
  {journal} {Science}\ }\textbf {\bibinfo {volume} {276}},\ \bibinfo {pages}
  {2016--2021} (\bibinfo {year} {1997})}\BibitemShut {NoStop}%
\bibitem [{\citenamefont {Kindt}\ \emph {et~al.}(2001)\citenamefont {Kindt},
  \citenamefont {Tzlil}, \citenamefont {Ben-Shaul},\ and\ \citenamefont
  {Gelbart}}]{kindt2001dna}%
  \BibitemOpen
  \bibfield  {author} {\bibinfo {author} {\bibfnamefont {J.}~\bibnamefont
  {Kindt}}, \bibinfo {author} {\bibfnamefont {S.}~\bibnamefont {Tzlil}},
  \bibinfo {author} {\bibfnamefont {A.}~\bibnamefont {Ben-Shaul}}, \ and\
  \bibinfo {author} {\bibfnamefont {W.M.}\ \bibnamefont {Gelbart}},\ }\bibfield
   {title} {\enquote {\bibinfo {title} {{DNA} packaging and ejection forces in
  bacteriophage},}\ }\href@noop {} {\bibfield  {journal} {\bibinfo  {journal}
  {Proc. Natl. Acad. Sci. U.S.A.}\ }\textbf {\bibinfo {volume} {98}},\ \bibinfo
  {pages} {13671--13674} (\bibinfo {year} {2001})}\BibitemShut {NoStop}%
\bibitem [{\citenamefont {Purohit}\ \emph {et~al.}(2003)\citenamefont
  {Purohit}, \citenamefont {Kondev},\ and\ \citenamefont
  {Phillips}}]{purohit2003}%
  \BibitemOpen
  \bibfield  {author} {\bibinfo {author} {\bibfnamefont {P.K.}\ \bibnamefont
  {Purohit}}, \bibinfo {author} {\bibfnamefont {J.}~\bibnamefont {Kondev}}, \
  and\ \bibinfo {author} {\bibfnamefont {R.}~\bibnamefont {Phillips}},\
  }\bibfield  {title} {\enquote {\bibinfo {title} {Mechanics of {DNA} packaging
  in viruses},}\ }\href@noop {} {\bibfield  {journal} {\bibinfo  {journal}
  {Proc. Natl. Acad. Sci. U.S.A.}\ }\textbf {\bibinfo {volume} {100}},\
  \bibinfo {pages} {3173--3178} (\bibinfo {year} {2003})}\BibitemShut {NoStop}%
\bibitem [{\citenamefont {Lu}\ and\ \citenamefont
  {Deutsch}(2005)}]{lu2005folding}%
  \BibitemOpen
  \bibfield  {author} {\bibinfo {author} {\bibfnamefont {J.}~\bibnamefont
  {Lu}}\ and\ \bibinfo {author} {\bibfnamefont {C.}~\bibnamefont {Deutsch}},\
  }\bibfield  {title} {\enquote {\bibinfo {title} {Folding zones inside the
  ribosomal exit tunnel},}\ }\href@noop {} {\bibfield  {journal} {\bibinfo
  {journal} {Nat. Struct. Mol. Biol.}\ }\textbf {\bibinfo {volume} {12}},\
  \bibinfo {pages} {1123--1129} (\bibinfo {year} {2005})}\BibitemShut {NoStop}%
\bibitem [{\citenamefont {Squires}\ and\ \citenamefont
  {Quake}(2005)}]{squires2005}%
  \BibitemOpen
  \bibfield  {author} {\bibinfo {author} {\bibfnamefont {T.M.}\ \bibnamefont
  {Squires}}\ and\ \bibinfo {author} {\bibfnamefont {S.R.}\ \bibnamefont
  {Quake}},\ }\bibfield  {title} {\enquote {\bibinfo {title} {Microfluidics:
  {F}luid physics at the nanoliter scale},}\ }\href@noop {} {\bibfield
  {journal} {\bibinfo  {journal} {Rev. Mod. Phys.}\ }\textbf {\bibinfo {volume}
  {77}},\ \bibinfo {pages} {977--1026} (\bibinfo {year} {2005})}\BibitemShut
  {NoStop}%
\bibitem [{\citenamefont {Jun}\ and\ \citenamefont
  {Mulder}(2006)}]{jun2006entropy}%
  \BibitemOpen
  \bibfield  {author} {\bibinfo {author} {\bibfnamefont {S.}~\bibnamefont
  {Jun}}\ and\ \bibinfo {author} {\bibfnamefont {B.}~\bibnamefont {Mulder}},\
  }\bibfield  {title} {\enquote {\bibinfo {title} {Entropy-driven spatial
  organization of highly confined polymers: {L}essons for the bacterial
  chromosome},}\ }\href@noop {} {\bibfield  {journal} {\bibinfo  {journal}
  {Proc. Nat. Acad. Sci. U.S.A.}\ }\textbf {\bibinfo {volume} {103}},\ \bibinfo
  {pages} {12388--12393} (\bibinfo {year} {2006})}\BibitemShut {NoStop}%
\bibitem [{\citenamefont {Persson}\ and\ \citenamefont
  {Tegenfeldt}(2010)}]{persson2010dna}%
  \BibitemOpen
  \bibfield  {author} {\bibinfo {author} {\bibfnamefont {F.}~\bibnamefont
  {Persson}}\ and\ \bibinfo {author} {\bibfnamefont {J.O.}\ \bibnamefont
  {Tegenfeldt}},\ }\bibfield  {title} {\enquote {\bibinfo {title} {{DNA} in
  nanochannels—directly visualizing genomic information},}\ }\href@noop {}
  {\bibfield  {journal} {\bibinfo  {journal} {Chem. Soc. Rev.}\ }\textbf
  {\bibinfo {volume} {39}},\ \bibinfo {pages} {985--999} (\bibinfo {year}
  {2010})}\BibitemShut {NoStop}%
\bibitem [{\citenamefont {Das}\ and\ \citenamefont
  {Chakraborty}(2010)}]{das2010effect}%
  \BibitemOpen
  \bibfield  {author} {\bibinfo {author} {\bibfnamefont {S.}~\bibnamefont
  {Das}}\ and\ \bibinfo {author} {\bibfnamefont {S.}~\bibnamefont
  {Chakraborty}},\ }\bibfield  {title} {\enquote {\bibinfo {title} {Effect of
  confinement on the collapsing mechanism of a flexible polymer chain},}\
  }\href@noop {} {\bibfield  {journal} {\bibinfo  {journal} {J. Chem. Phys.}\
  }\textbf {\bibinfo {volume} {133}} (\bibinfo {year} {2010})}\BibitemShut
  {NoStop}%
\bibitem [{\citenamefont {Majumder}\ and\ \citenamefont
  {Janke}(2016{\natexlab{b}})}]{majumder2016aging}%
  \BibitemOpen
  \bibfield  {author} {\bibinfo {author} {\bibfnamefont {S.}~\bibnamefont
  {Majumder}}\ and\ \bibinfo {author} {\bibfnamefont {W.}~\bibnamefont
  {Janke}},\ }\bibfield  {title} {\enquote {\bibinfo {title} {Aging and related
  scaling during the collapse of a polymer},}\ }\bibfield  {booktitle} {\emph
  {\bibinfo {booktitle} {J. Phys. Conf. Ser.}},\ }\href@noop {} {\ \textbf
  {\bibinfo {volume} {750}},\ \bibinfo {pages} {012020} (\bibinfo {year}
  {2016}{\natexlab{b}})}\BibitemShut {NoStop}%
\bibitem [{\citenamefont {Majumder}\ \emph
  {et~al.}(2018{\natexlab{a}})\citenamefont {Majumder}, \citenamefont
  {Christiansen},\ and\ \citenamefont {Janke}}]{majumder2018scaling}%
  \BibitemOpen
  \bibfield  {author} {\bibinfo {author} {\bibfnamefont {S.}~\bibnamefont
  {Majumder}}, \bibinfo {author} {\bibfnamefont {H.}~\bibnamefont
  {Christiansen}}, \ and\ \bibinfo {author} {\bibfnamefont {W.}~\bibnamefont
  {Janke}},\ }\bibfield  {title} {\enquote {\bibinfo {title} {Scaling laws
  during collapse of a homopolymer: {L}attice versus off-lattice},}\ }\bibfield
   {booktitle} {\emph {\bibinfo {booktitle} {J. Phys. Conf. Ser.}},\
  }\href@noop {} {\ \textbf {\bibinfo {volume} {955}},\ \bibinfo {pages}
  {012008} (\bibinfo {year} {2018}{\natexlab{a}})}\BibitemShut {NoStop}%
\bibitem [{\citenamefont {Klushin}(1998)}]{klushin1998}%
  \BibitemOpen
  \bibfield  {author} {\bibinfo {author} {\bibfnamefont {L.I.}\ \bibnamefont
  {Klushin}},\ }\bibfield  {title} {\enquote {\bibinfo {title} {Kinetics of a
  homopolymer collapse: {Beyond the Rouse--Zimm scaling}},}\ }\href@noop {}
  {\bibfield  {journal} {\bibinfo  {journal} {J. Chem. Phys.}\ }\textbf
  {\bibinfo {volume} {108}},\ \bibinfo {pages} {7917--7920} (\bibinfo {year}
  {1998})}\BibitemShut {NoStop}%
\bibitem [{\citenamefont {Kremer}\ and\ \citenamefont
  {Grest}(1990)}]{kremer1990dynamics}%
  \BibitemOpen
  \bibfield  {author} {\bibinfo {author} {\bibfnamefont {K.}~\bibnamefont
  {Kremer}}\ and\ \bibinfo {author} {\bibfnamefont {G.~S.}\ \bibnamefont
  {Grest}},\ }\bibfield  {title} {\enquote {\bibinfo {title} {Dynamics of
  entangled linear polymer melts: {A} molecular-dynamics simulation},}\
  }\href@noop {} {\bibfield  {journal} {\bibinfo  {journal} {J. Chem. Phys.}\
  }\textbf {\bibinfo {volume} {92}},\ \bibinfo {pages} {5057--5086} (\bibinfo
  {year} {1990})}\BibitemShut {NoStop}%
\bibitem [{\citenamefont {Frenkel}\ and\ \citenamefont {Smit}(2002)}]{frenkel}%
  \BibitemOpen
  \bibfield  {author} {\bibinfo {author} {\bibfnamefont {D.}~\bibnamefont
  {Frenkel}}\ and\ \bibinfo {author} {\bibfnamefont {B.}~\bibnamefont {Smit}},\
  }\href@noop {} {\emph {\bibinfo {title} {Understanding Molecular Simulations:
  From Algorithms to Applications}}}\ (\bibinfo  {publisher} {San Diego:
  Academic Press},\ \bibinfo {year} {2002})\BibitemShut {NoStop}%
\bibitem [{\citenamefont {Nos{\'e}}(1984)}]{nose1984unified}%
  \BibitemOpen
  \bibfield  {author} {\bibinfo {author} {\bibfnamefont {S.}~\bibnamefont
  {Nos{\'e}}},\ }\bibfield  {title} {\enquote {\bibinfo {title} {A unified
  formulation of the constant temperature molecular dynamics methods},}\
  }\href@noop {} {\bibfield  {journal} {\bibinfo  {journal} {J. Chem. Phys.}\
  }\textbf {\bibinfo {volume} {81}},\ \bibinfo {pages} {511--519} (\bibinfo
  {year} {1984})}\BibitemShut {NoStop}%
\bibitem [{\citenamefont {Hoover}(1985)}]{hoover1985canonical}%
  \BibitemOpen
  \bibfield  {author} {\bibinfo {author} {\bibfnamefont {W.~G.}\ \bibnamefont
  {Hoover}},\ }\bibfield  {title} {\enquote {\bibinfo {title} {Canonical
  dynamics: Equilibrium phase-space distributions},}\ }\href@noop {} {\bibfield
   {journal} {\bibinfo  {journal} {Phys. Rev. A}\ }\textbf {\bibinfo {volume}
  {31}},\ \bibinfo {pages} {1695} (\bibinfo {year} {1985})}\BibitemShut
  {NoStop}%
\bibitem [{\citenamefont {Martyna}\ \emph {et~al.}(1992)\citenamefont
  {Martyna}, \citenamefont {Klein},\ and\ \citenamefont
  {Tuckerman}}]{martyna1992nose}%
  \BibitemOpen
  \bibfield  {author} {\bibinfo {author} {\bibfnamefont {G.~J.}\ \bibnamefont
  {Martyna}}, \bibinfo {author} {\bibfnamefont {M.~L.}\ \bibnamefont {Klein}},
  \ and\ \bibinfo {author} {\bibfnamefont {M.}~\bibnamefont {Tuckerman}},\
  }\bibfield  {title} {\enquote {\bibinfo {title} {{Nos{\'e}--Hoover} chains:
  {T}he canonical ensemble via continuous dynamics},}\ }\href@noop {}
  {\bibfield  {journal} {\bibinfo  {journal} {J. Chem. Phys.}\ }\textbf
  {\bibinfo {volume} {97}},\ \bibinfo {pages} {2635--2643} (\bibinfo {year}
  {1992})}\BibitemShut {NoStop}%
\bibitem [{\citenamefont {Majumder}\ and\ \citenamefont
  {Das}(2013{\natexlab{a}})}]{majumder2013effects}%
  \BibitemOpen
  \bibfield  {author} {\bibinfo {author} {\bibfnamefont {S.}~\bibnamefont
  {Majumder}}\ and\ \bibinfo {author} {\bibfnamefont {S.K.}\ \bibnamefont
  {Das}},\ }\bibfield  {title} {\enquote {\bibinfo {title} {Effects of density
  conservation and hydrodynamics on aging in nonequilibrium processes},}\
  }\href@noop {} {\bibfield  {journal} {\bibinfo  {journal} {Phys. Rev. Lett.}\
  }\textbf {\bibinfo {volume} {111}},\ \bibinfo {pages} {055503} (\bibinfo
  {year} {2013}{\natexlab{a}})}\BibitemShut {NoStop}%
\bibitem [{\citenamefont {Plimpton}(1995)}]{plimpton1995}%
  \BibitemOpen
  \bibfield  {author} {\bibinfo {author} {\bibfnamefont {S.}~\bibnamefont
  {Plimpton}},\ }\bibfield  {title} {\enquote {\bibinfo {title} {Fast parallel
  algorithms for short-range molecular dynamics},}\ }\href@noop {} {\bibfield
  {journal} {\bibinfo  {journal} {J. Comput. Phys.}\ }\textbf {\bibinfo
  {volume} {117}},\ \bibinfo {pages} {1--19} (\bibinfo {year}
  {1995})}\BibitemShut {NoStop}%
\bibitem [{\citenamefont {Scalley}\ and\ \citenamefont
  {Baker}(1997)}]{scalley1997}%
  \BibitemOpen
  \bibfield  {author} {\bibinfo {author} {\bibfnamefont {M.L.}\ \bibnamefont
  {Scalley}}\ and\ \bibinfo {author} {\bibfnamefont {D.}~\bibnamefont
  {Baker}},\ }\bibfield  {title} {\enquote {\bibinfo {title} {Protein folding
  kinetics exhibit an {A}rrhenius temperature dependence when corrected for the
  temperature dependence of protein stability},}\ }\href@noop {} {\bibfield
  {journal} {\bibinfo  {journal} {Proc. Natl. Acad. Sci. U.S.A.}\ }\textbf
  {\bibinfo {volume} {94}},\ \bibinfo {pages} {10636--10640} (\bibinfo {year}
  {1997})}\BibitemShut {NoStop}%
\bibitem [{\citenamefont {Eaton}\ \emph {et~al.}(2000)\citenamefont {Eaton},
  \citenamefont {Munoz}, \citenamefont {Hagen}, \citenamefont {Jas},
  \citenamefont {Lapidus}, \citenamefont {Henry},\ and\ \citenamefont
  {Hofrichter}}]{eaton2000fast}%
  \BibitemOpen
  \bibfield  {author} {\bibinfo {author} {\bibfnamefont {W.A.}\ \bibnamefont
  {Eaton}}, \bibinfo {author} {\bibfnamefont {V.}~\bibnamefont {Munoz}},
  \bibinfo {author} {\bibfnamefont {S.J.}\ \bibnamefont {Hagen}}, \bibinfo
  {author} {\bibfnamefont {G.S.}\ \bibnamefont {Jas}}, \bibinfo {author}
  {\bibfnamefont {L.J.}\ \bibnamefont {Lapidus}}, \bibinfo {author}
  {\bibfnamefont {E.R.}\ \bibnamefont {Henry}}, \ and\ \bibinfo {author}
  {\bibfnamefont {J.}~\bibnamefont {Hofrichter}},\ }\bibfield  {title}
  {\enquote {\bibinfo {title} {Fast kinetics and mechanisms in protein
  folding},}\ }\href@noop {} {\bibfield  {journal} {\bibinfo  {journal} {Annu.
  Rev. Biophys. Biomol. Struct.}\ }\textbf {\bibinfo {volume} {29}},\ \bibinfo
  {pages} {327--359} (\bibinfo {year} {2000})}\BibitemShut {NoStop}%
\bibitem [{\citenamefont {Naganathan}\ \emph {et~al.}(2007)\citenamefont
  {Naganathan}, \citenamefont {Doshi},\ and\ \citenamefont
  {Mu{\~n}oz}}]{naganathan2007protein}%
  \BibitemOpen
  \bibfield  {author} {\bibinfo {author} {\bibfnamefont {A.N.}\ \bibnamefont
  {Naganathan}}, \bibinfo {author} {\bibfnamefont {U.}~\bibnamefont {Doshi}}, \
  and\ \bibinfo {author} {\bibfnamefont {V.}~\bibnamefont {Mu{\~n}oz}},\
  }\bibfield  {title} {\enquote {\bibinfo {title} {Protein folding kinetics:
  {B}arrier effects in chemical and thermal denaturation experiments},}\
  }\href@noop {} {\bibfield  {journal} {\bibinfo  {journal} {J. Am. Chem.
  Soc.}\ }\textbf {\bibinfo {volume} {129}},\ \bibinfo {pages} {5673--5682}
  (\bibinfo {year} {2007})}\BibitemShut {NoStop}%
\bibitem [{\citenamefont {Majumder}\ and\ \citenamefont
  {Das}(2013{\natexlab{b}})}]{majumder2013temperature}%
  \BibitemOpen
  \bibfield  {author} {\bibinfo {author} {\bibfnamefont {S.}~\bibnamefont
  {Majumder}}\ and\ \bibinfo {author} {\bibfnamefont {S.K.}\ \bibnamefont
  {Das}},\ }\bibfield  {title} {\enquote {\bibinfo {title} {Temperature and
  composition dependence of kinetics of phase separation in solid binary
  mixtures},}\ }\href@noop {} {\bibfield  {journal} {\bibinfo  {journal} {Phys.
  Chem. Chem. Phys.}\ }\textbf {\bibinfo {volume} {15}},\ \bibinfo {pages}
  {13209--13218} (\bibinfo {year} {2013}{\natexlab{b}})}\BibitemShut {NoStop}%
\bibitem [{\citenamefont {Majumder}\ \emph
  {et~al.}(2018{\natexlab{b}})\citenamefont {Majumder}, \citenamefont {Das},\
  and\ \citenamefont {Janke}}]{majumder2018universal}%
  \BibitemOpen
  \bibfield  {author} {\bibinfo {author} {\bibfnamefont {S.}~\bibnamefont
  {Majumder}}, \bibinfo {author} {\bibfnamefont {S.K.}\ \bibnamefont {Das}}, \
  and\ \bibinfo {author} {\bibfnamefont {W.}~\bibnamefont {Janke}},\ }\bibfield
   {title} {\enquote {\bibinfo {title} {Universal finite-size scaling function
  for kinetics of phase separation in mixtures with varying number of
  components},}\ }\href@noop {} {\bibfield  {journal} {\bibinfo  {journal}
  {Phys. Rev. E}\ }\textbf {\bibinfo {volume} {98}},\ \bibinfo {pages} {042142}
  (\bibinfo {year} {2018}{\natexlab{b}})}\BibitemShut {NoStop}%
\bibitem [{\citenamefont {Basu}\ \emph {et~al.}(2016)\citenamefont {Basu},
  \citenamefont {Majumder}, \citenamefont {Sutradhar}, \citenamefont {Das},\
  and\ \citenamefont {Paul}}]{basu2016phase}%
  \BibitemOpen
  \bibfield  {author} {\bibinfo {author} {\bibfnamefont {S.}~\bibnamefont
  {Basu}}, \bibinfo {author} {\bibfnamefont {S.}~\bibnamefont {Majumder}},
  \bibinfo {author} {\bibfnamefont {S.}~\bibnamefont {Sutradhar}}, \bibinfo
  {author} {\bibfnamefont {S.K.}\ \bibnamefont {Das}}, \ and\ \bibinfo {author}
  {\bibfnamefont {R.}~\bibnamefont {Paul}},\ }\bibfield  {title} {\enquote
  {\bibinfo {title} {Phase segregation in a binary fluid confined inside a
  nanopore},}\ }\href@noop {} {\bibfield  {journal} {\bibinfo  {journal}
  {Europhys. Lett.}\ }\textbf {\bibinfo {volume} {116}},\ \bibinfo {pages}
  {56003} (\bibinfo {year} {2016})}\BibitemShut {NoStop}%
\bibitem [{\citenamefont {Basu}\ \emph {et~al.}(2026)\citenamefont {Basu},
  \citenamefont {Majumder}, \citenamefont {Paul},\ and\ \citenamefont
  {Das}}]{basu2026}%
  \BibitemOpen
  \bibfield  {author} {\bibinfo {author} {\bibfnamefont {S.}~\bibnamefont
  {Basu}}, \bibinfo {author} {\bibfnamefont {S.}~\bibnamefont {Majumder}},
  \bibinfo {author} {\bibfnamefont {R.}~\bibnamefont {Paul}}, \ and\ \bibinfo
  {author} {\bibfnamefont {S.K.}\ \bibnamefont {Das}},\ }\bibfield  {title}
  {\enquote {\bibinfo {title} {Domain growth and aging in a phase separating
  binary fluid confined inside a nanopore},}\ }\href@noop {} {\bibfield
  {journal} {\bibinfo  {journal} {Soft Matter}\ }\textbf {\bibinfo {volume}
  {22}},\ \bibinfo {pages} {1251--1261} (\bibinfo {year} {2026})}\BibitemShut
  {NoStop}%
\bibitem [{\citenamefont {Tegenfeldt}\ \emph {et~al.}(2004)\citenamefont
  {Tegenfeldt}, \citenamefont {Prinz}, \citenamefont {Cao}, \citenamefont
  {Chou}, \citenamefont {Reisner}, \citenamefont {Riehn}, \citenamefont {Wang},
  \citenamefont {Cox}, \citenamefont {Sturm}, \citenamefont {Silberzan} \emph
  {et~al.}}]{tegenfeldt2004dynamics}%
  \BibitemOpen
  \bibfield  {author} {\bibinfo {author} {\bibfnamefont {J.O.}\ \bibnamefont
  {Tegenfeldt}}, \bibinfo {author} {\bibfnamefont {C.}~\bibnamefont {Prinz}},
  \bibinfo {author} {\bibfnamefont {H.}~\bibnamefont {Cao}}, \bibinfo {author}
  {\bibfnamefont {S.}~\bibnamefont {Chou}}, \bibinfo {author} {\bibfnamefont
  {W.W.}\ \bibnamefont {Reisner}}, \bibinfo {author} {\bibfnamefont
  {R.}~\bibnamefont {Riehn}}, \bibinfo {author} {\bibfnamefont {Y.M.}\
  \bibnamefont {Wang}}, \bibinfo {author} {\bibfnamefont {E.C.}\ \bibnamefont
  {Cox}}, \bibinfo {author} {\bibfnamefont {J.C.}\ \bibnamefont {Sturm}},
  \bibinfo {author} {\bibfnamefont {P.}~\bibnamefont {Silberzan}},  \emph
  {et~al.},\ }\bibfield  {title} {\enquote {\bibinfo {title} {The dynamics of
  genomic-length {DNA} molecules in 100-nm channels},}\ }\href@noop {}
  {\bibfield  {journal} {\bibinfo  {journal} {Proc. Natl. Acad. Sci. U.S.A.}\
  }\textbf {\bibinfo {volume} {101}},\ \bibinfo {pages} {10979--10983}
  (\bibinfo {year} {2004})}\BibitemShut {NoStop}%
\bibitem [{\citenamefont {Reisner}\ \emph {et~al.}(2005)\citenamefont
  {Reisner}, \citenamefont {Morton}, \citenamefont {Riehn}, \citenamefont
  {Wang}, \citenamefont {Yu}, \citenamefont {Rosen}, \citenamefont {Sturm},
  \citenamefont {Chou}, \citenamefont {Frey},\ and\ \citenamefont
  {Austin}}]{reisner2005statics}%
  \BibitemOpen
  \bibfield  {author} {\bibinfo {author} {\bibfnamefont {W.}~\bibnamefont
  {Reisner}}, \bibinfo {author} {\bibfnamefont {K.J.}\ \bibnamefont {Morton}},
  \bibinfo {author} {\bibfnamefont {R.}~\bibnamefont {Riehn}}, \bibinfo
  {author} {\bibfnamefont {Y.M.}\ \bibnamefont {Wang}}, \bibinfo {author}
  {\bibfnamefont {Z.}~\bibnamefont {Yu}}, \bibinfo {author} {\bibfnamefont
  {M.}~\bibnamefont {Rosen}}, \bibinfo {author} {\bibfnamefont {J.C.}\
  \bibnamefont {Sturm}}, \bibinfo {author} {\bibfnamefont {S.Y.}\ \bibnamefont
  {Chou}}, \bibinfo {author} {\bibfnamefont {E.}~\bibnamefont {Frey}}, \ and\
  \bibinfo {author} {\bibfnamefont {R.H.}\ \bibnamefont {Austin}},\ }\bibfield
  {title} {\enquote {\bibinfo {title} {Statics and dynamics of single {DNA}
  molecules confined in nanochannels},}\ }\href@noop {} {\bibfield  {journal}
  {\bibinfo  {journal} {Phys. Rev. Lett.}\ }\textbf {\bibinfo {volume} {94}},\
  \bibinfo {pages} {196101} (\bibinfo {year} {2005})}\BibitemShut {NoStop}%
\bibitem [{\citenamefont {Reisner}\ \emph {et~al.}(2007)\citenamefont
  {Reisner}, \citenamefont {Beech}, \citenamefont {Larsen}, \citenamefont
  {Flyvbjerg}, \citenamefont {Kristensen},\ and\ \citenamefont
  {Tegenfeldt}}]{reisner2007nanoconfinement}%
  \BibitemOpen
  \bibfield  {author} {\bibinfo {author} {\bibfnamefont {W.W.}\ \bibnamefont
  {Reisner}}, \bibinfo {author} {\bibfnamefont {J.P.}\ \bibnamefont {Beech}},
  \bibinfo {author} {\bibfnamefont {N.B.}\ \bibnamefont {Larsen}}, \bibinfo
  {author} {\bibfnamefont {H.}~\bibnamefont {Flyvbjerg}}, \bibinfo {author}
  {\bibfnamefont {A.}~\bibnamefont {Kristensen}}, \ and\ \bibinfo {author}
  {\bibfnamefont {J.O.}\ \bibnamefont {Tegenfeldt}},\ }\bibfield  {title}
  {\enquote {\bibinfo {title} {Nanoconfinement-enhanced conformational response
  of single {DNA} molecules to changes in ionic environment},}\ }\href@noop {}
  {\bibfield  {journal} {\bibinfo  {journal} {Phys. Rev. Lett.}\ }\textbf
  {\bibinfo {volume} {99}},\ \bibinfo {pages} {058302} (\bibinfo {year}
  {2007})}\BibitemShut {NoStop}%
\bibitem [{\citenamefont {Hofmann}\ \emph {et~al.}(2012)\citenamefont
  {Hofmann}, \citenamefont {Soranno}, \citenamefont {Borgia}, \citenamefont
  {Gast}, \citenamefont {Nettels},\ and\ \citenamefont
  {Schuler}}]{hofmann2012polymer}%
  \BibitemOpen
  \bibfield  {author} {\bibinfo {author} {\bibfnamefont {H.}~\bibnamefont
  {Hofmann}}, \bibinfo {author} {\bibfnamefont {A.}~\bibnamefont {Soranno}},
  \bibinfo {author} {\bibfnamefont {A.}~\bibnamefont {Borgia}}, \bibinfo
  {author} {\bibfnamefont {K.}~\bibnamefont {Gast}}, \bibinfo {author}
  {\bibfnamefont {D.}~\bibnamefont {Nettels}}, \ and\ \bibinfo {author}
  {\bibfnamefont {B.}~\bibnamefont {Schuler}},\ }\bibfield  {title} {\enquote
  {\bibinfo {title} {Polymer scaling laws of unfolded and intrinsically
  disordered proteins quantified with single-molecule spectroscopy},}\
  }\href@noop {} {\bibfield  {journal} {\bibinfo  {journal} {Proc. Natl. Acad.
  Sci. U.S.A.}\ }\textbf {\bibinfo {volume} {109}},\ \bibinfo {pages}
  {16155--16160} (\bibinfo {year} {2012})}\BibitemShut {NoStop}%
\bibitem [{\citenamefont {Tress}\ \emph {et~al.}(2013)\citenamefont {Tress},
  \citenamefont {Mapesa}, \citenamefont {Kossack}, \citenamefont {Kipnusu},
  \citenamefont {Reiche},\ and\ \citenamefont {Kremer}}]{tress2013glassy}%
  \BibitemOpen
  \bibfield  {author} {\bibinfo {author} {\bibfnamefont {M.}~\bibnamefont
  {Tress}}, \bibinfo {author} {\bibfnamefont {E.U.}\ \bibnamefont {Mapesa}},
  \bibinfo {author} {\bibfnamefont {W.}~\bibnamefont {Kossack}}, \bibinfo
  {author} {\bibfnamefont {W.K.}\ \bibnamefont {Kipnusu}}, \bibinfo {author}
  {\bibfnamefont {M.}~\bibnamefont {Reiche}}, \ and\ \bibinfo {author}
  {\bibfnamefont {F.}~\bibnamefont {Kremer}},\ }\bibfield  {title} {\enquote
  {\bibinfo {title} {Glassy dynamics in condensed isolated polymer chains},}\
  }\href@noop {} {\bibfield  {journal} {\bibinfo  {journal} {Science}\ }\textbf
  {\bibinfo {volume} {341}},\ \bibinfo {pages} {1371--1374} (\bibinfo {year}
  {2013})}\BibitemShut {NoStop}%
\bibitem [{\citenamefont {Besteman}\ \emph {et~al.}(2007)\citenamefont
  {Besteman}, \citenamefont {Hage}, \citenamefont {Dekker},\ and\ \citenamefont
  {Lemay}}]{besteman2007role}%
  \BibitemOpen
  \bibfield  {author} {\bibinfo {author} {\bibfnamefont {K.}~\bibnamefont
  {Besteman}}, \bibinfo {author} {\bibfnamefont {S.}~\bibnamefont {Hage}},
  \bibinfo {author} {\bibfnamefont {N.H.}\ \bibnamefont {Dekker}}, \ and\
  \bibinfo {author} {\bibfnamefont {S.G.}\ \bibnamefont {Lemay}},\ }\bibfield
  {title} {\enquote {\bibinfo {title} {Role of tension and twist in
  single-molecule {DNA} condensation},}\ }\href@noop {} {\bibfield  {journal}
  {\bibinfo  {journal} {Phys. Rev. Lett.}\ }\textbf {\bibinfo {volume} {98}},\
  \bibinfo {pages} {058103} (\bibinfo {year} {2007})}\BibitemShut {NoStop}%
\bibitem [{\citenamefont {Adelsberger}\ \emph {et~al.}(2012)\citenamefont
  {Adelsberger}, \citenamefont {Metwalli}, \citenamefont {Diethert},
  \citenamefont {Grillo}, \citenamefont {Bivigou-Koumba}, \citenamefont
  {Laschewsky}, \citenamefont {M{\"u}ller-Buschbaum},\ and\ \citenamefont
  {Papadakis}}]{adelsberger2012kinetics}%
  \BibitemOpen
  \bibfield  {author} {\bibinfo {author} {\bibfnamefont {J.}~\bibnamefont
  {Adelsberger}}, \bibinfo {author} {\bibfnamefont {E.}~\bibnamefont
  {Metwalli}}, \bibinfo {author} {\bibfnamefont {A.}~\bibnamefont {Diethert}},
  \bibinfo {author} {\bibfnamefont {I.}~\bibnamefont {Grillo}}, \bibinfo
  {author} {\bibfnamefont {A.M.}\ \bibnamefont {Bivigou-Koumba}}, \bibinfo
  {author} {\bibfnamefont {A.}~\bibnamefont {Laschewsky}}, \bibinfo {author}
  {\bibfnamefont {P.}~\bibnamefont {M{\"u}ller-Buschbaum}}, \ and\ \bibinfo
  {author} {\bibfnamefont {C.M.}\ \bibnamefont {Papadakis}},\ }\bibfield
  {title} {\enquote {\bibinfo {title} {Kinetics of collapse transition and
  cluster formation in a thermoresponsive micellar solution of {P
  (S-b-NIPAM-b-S)} induced by a temperature jump},}\ }\href@noop {} {\bibfield
  {journal} {\bibinfo  {journal} {Macromol. Rapid Commun.}\ }\textbf {\bibinfo
  {volume} {33}},\ \bibinfo {pages} {254--259} (\bibinfo {year}
  {2012})}\BibitemShut {NoStop}%
\bibitem [{\citenamefont {Alberts}\ \emph {et~al.}(1994)\citenamefont
  {Alberts}, \citenamefont {Bray}, \citenamefont {Lewis}, \citenamefont {Raff},
  \citenamefont {Roberts}, \citenamefont {Watson} \emph
  {et~al.}}]{alberts1994molecular}%
  \BibitemOpen
  \bibfield  {author} {\bibinfo {author} {\bibfnamefont {B.}~\bibnamefont
  {Alberts}}, \bibinfo {author} {\bibfnamefont {D.}~\bibnamefont {Bray}},
  \bibinfo {author} {\bibfnamefont {J.}~\bibnamefont {Lewis}}, \bibinfo
  {author} {\bibfnamefont {M.}~\bibnamefont {Raff}}, \bibinfo {author}
  {\bibfnamefont {K.}~\bibnamefont {Roberts}}, \bibinfo {author} {\bibfnamefont
  {J.D.}\ \bibnamefont {Watson}},  \emph {et~al.},\ }\href@noop {} {\emph
  {\bibinfo {title} {Molecular Biology of the Cell}}},\ Vol.~\bibinfo {volume}
  {3}\ (\bibinfo  {publisher} {Garland New York},\ \bibinfo {year}
  {1994})\BibitemShut {NoStop}%
\bibitem [{\citenamefont {Packer}(1963)}]{packer1963size}%
  \BibitemOpen
  \bibfield  {author} {\bibinfo {author} {\bibfnamefont {L.}~\bibnamefont
  {Packer}},\ }\bibfield  {title} {\enquote {\bibinfo {title} {Size and shape
  transformations correlated with oxidative phosphorylation in mitochondria: I.
  swelling-shrinkage mechanisms in intact mitochondria},}\ }\href@noop {}
  {\bibfield  {journal} {\bibinfo  {journal} {J. Cell. Biol.}\ }\textbf
  {\bibinfo {volume} {18}},\ \bibinfo {pages} {487--494} (\bibinfo {year}
  {1963})}\BibitemShut {NoStop}%
\bibitem [{\citenamefont {Meyers}\ \emph {et~al.}(2006)\citenamefont {Meyers},
  \citenamefont {Craig},\ and\ \citenamefont {Odde}}]{meyers2006potential}%
  \BibitemOpen
  \bibfield  {author} {\bibinfo {author} {\bibfnamefont {J.}~\bibnamefont
  {Meyers}}, \bibinfo {author} {\bibfnamefont {J.}~\bibnamefont {Craig}}, \
  and\ \bibinfo {author} {\bibfnamefont {D.J.}\ \bibnamefont {Odde}},\
  }\bibfield  {title} {\enquote {\bibinfo {title} {Potential for control of
  signaling pathways via cell size and shape},}\ }\href@noop {} {\bibfield
  {journal} {\bibinfo  {journal} {Curr. Biol.}\ }\textbf {\bibinfo {volume}
  {16}},\ \bibinfo {pages} {1685--1693} (\bibinfo {year} {2006})}\BibitemShut
  {NoStop}%
\bibitem [{\citenamefont {Lizana}\ \emph {et~al.}(2008)\citenamefont {Lizana},
  \citenamefont {Bauer},\ and\ \citenamefont {Orwar}}]{lizana2008controlling}%
  \BibitemOpen
  \bibfield  {author} {\bibinfo {author} {\bibfnamefont {L.}~\bibnamefont
  {Lizana}}, \bibinfo {author} {\bibfnamefont {B.}~\bibnamefont {Bauer}}, \
  and\ \bibinfo {author} {\bibfnamefont {O.}~\bibnamefont {Orwar}},\ }\bibfield
   {title} {\enquote {\bibinfo {title} {Controlling the rates of biochemical
  reactions and signaling networks by shape and volume changes},}\ }\href@noop
  {} {\bibfield  {journal} {\bibinfo  {journal} {Proc. Natl. Acad. Sci. U.S.A}\
  }\textbf {\bibinfo {volume} {105}},\ \bibinfo {pages} {4099--4104} (\bibinfo
  {year} {2008})}\BibitemShut {NoStop}%
\end{thebibliography}
%
\end{document}